\documentclass{aa}

\usepackage{graphicx}
\usepackage{dcolumn}
\usepackage{bm}

\usepackage{txfonts}
\usepackage{natbib}
\usepackage{amsmath}    
\usepackage{graphicx}   
\usepackage{verbatim}   
\usepackage{color}      
\usepackage{subfigure}  
\usepackage{gensymb} 
\usepackage{wasysym} 
\usepackage{hyperref}   
\bibpunct{(}{)}{;}{a}{}{,} 
\usepackage{graphicx}
\usepackage{nth}
\newcommand{\kms}{\mbox{${\rm km\,s}^{-1}$}}
\newcommand{\NaX}{\mbox{${\rm X_{\mathrm{NaP}}}$}}

\begin{document}

\title{Wind of Change: retrieving exoplanet atmospheric winds from high-resolution spectroscopy}
    
\author{J.~V.~Seidel\inst{1} 
\and D.~Ehrenreich\inst{1}
\and L.~Pino\inst{2}
\and V.~Bourrier\inst{1}
\and B.~Lavie\inst{1}
\and R.~Allart\inst{1}
\and A.~Wyttenbach\inst{3}
\and C.~Lovis\inst{1}
}

\institute{Observatoire astronomique de l'Universit\'e de Gen\`eve, chemin des Maillettes 51, 1290 Versoix, Switzerland
\and Anton Pannekoek Institute for Astronomy, University of Amsterdam, Science Park 904, 1098 XH Amsterdam, The Netherlands
\and Leiden Observatory, Leiden University, Postbus 9513, 2300 RA Leiden, The Netherlands
}

\date{Received date/ Accepted date}

\abstract{\textit{Context.} The atmosphere of exoplanets has been studied extensively in recent years, using numerical models to retrieve chemical composition, dynamical circulation or temperature from data. One of the best observational probes in transmission is the sodium doublet, due to its large cross section. However, modelling the shape of the planetary sodium lines has proven to be challenging. Models with different assumptions regarding the atmosphere have been employed to fit the lines in the literature, yet statistically sound direct comparisons of different models are needed to paint a clear picture. \\ \textit{Aims.} We will compare different wind and temperature patterns and provide a tool to distinguish them driven by their best fit for the sodium transmission spectrum of the hot Jupiter HD~189733b. We parametrise different possible wind patterns already tested in literature and introduce the new option of an upwards driven vertical wind. \\ \textit{Methods.} We construct a forward model where the wind speed, wind geometry and temperature are injected into the calculation of the transmission spectrum. We embed this forward model in a nested sampling retrieval code to rank the models via their Bayesian evidence. \\ \textit{Results.} We retrieve a best-fit to the HD~189733b data for vertical upward winds $|\vec{v}_{\mathrm{ver}}(\mathrm{mean})|=40\pm4~\kms$ at altitudes above $10^{-6}$~bar. With the current data from HARPS, we cannot distinguish wind patterns for higher pressure atmospheric layers.  \\ \textit{Conclusions.} We show that vertical upwards winds in the upper atmosphere are a possible explanation for the broad sodium signature in hot Jupiters. We highlight other influences on the width of the doublet and explore strong magnetic fields acting on the lower atmosphere as one possible origin of the retrieved wind speed.}

\keywords{Planetary Systems -- Planets and satellites: atmospheres, individual: HD~189733b -- Techniques: spectroscopic -- Line: profiles -- Methods: data analysis}
\titlerunning{Wind of Change: retrieving the atmospheric wind profile of exoplanets}
\maketitle

\section{Introduction}
Providing context to our own Solar System and the diverse atmospheres found therein is one of the goals of exoplanet research. While we have detailed constraints on the gas giants in our own Solar System thanks to the Galileo \citep{Mahaffy2000}, Cassini \citep{Lebreton1992}, and Juno \citep{Bolton2010} missions, it is just now that we have the precision to measure exoplanetary atmospheres and the range of confirmed exoplanets to start a more in depth discussion on atmospheric compositions \citep{Fletcher2019}. 
From hot giant planets like Jupiter down to Earth sized exoplanets, the atmosphere is accessible as a thin ring via remote transit observations among other methods. With these observations, a transmission spectrum can be built by taking the ratio of spectra taken during the transit and out of transit. The high-resolution transmission spectrum, showing the spectral lines of the planet atmosphere resolved individually, can be used to find new elements in the atmosphere and discuss different chemical compositions and dynamic patterns. 

The sodium doublet is well suited for this endeavour because it probes the atmosphere up to the thermosphere, thus allowing to study many orders of magnitude in pressure. The opportunity to use sodium as a high-altitude probe arises from its characteristic resonance doublet at 589 nm \citep{Seager2000}.

For hot Jupiters, a plethora of sodium detections have been confirmed, starting with \cite{Charbonneau2002} in low resolution and \cite{Snellen2008,Redfield2008} up to more recent high resolution observations (e. g. \cite{Louden2015,RiddenHarper2016, Wyttenbach2015, Wyttenbach2017,Casasayas-Barris2017,Jensen2018, Khalafinejad2018,Seidel2019}).

These high-resolution sodium detections show significantly broader lines than a static model of the atmosphere including pressure and temperature broadening would suggest. In the case of WASP~76b, the detected sodium doublet is ten times broader than the instrument response function \citep{Seidel2019}. Both WASP~49b and HD~189733b show similar broad features \citep{Wyttenbach2017, Casasayas-Barris2017}. In this study we will focus on the sodium feature of HD~189733b. To observe the sodium doublet with a high enough precision to resolve the line core, we used the HARPS (High-Accuracy Radial-velocity Planet Searcher) spectrograph at ESO's 3.6m telescope in La Silla, Chile, with a resolution of $R = 115~000$ \citep{Mayor2003}. 
A possible explanation for this broadening is Doppler broadening from strong winds in the atmosphere, as has been claimed for CO lines which probe a different altitude in the atmosphere \citep{Brogi2016}. Theoretical studies have suggested different wind patterns in the lower atmosphere, for example: super-rotational wind, where the atmosphere rotates faster than the planet itself (e. g. \cite{Showman2011}); day-to-night side wind, where the wind on a tidally locked planet is upwelling on the dayside and downwelling on the nightside (e. g. \cite{Flowers2019}); and lastly vertical wind patterns pushing the atmosphere away from the planet surface (e. g. \cite{Komacek2019,Debrecht2019}).  

Global circulation models (GCM) build a three dimensional atmosphere and are capable of calculating model atmospheres with equatorial jets and varying temperature profiles in latitude, longitude and altitude \citep{Showman2018,Flowers2019,Steinrueck2019,Deibert2019}. This level of sophistication means in almost all cases a high computational cost, limiting the possible number of executions. Especially for retrieval approaches, which require executions in the order of $100,000$, full GCM models are challenging. In this work, we introduce a forward model for transmission spectroscopy including wind patterns that operates on a one dimensional grid. The simplification of a 1D calculation allows to link the forward model to a multi nested sampling retrieval followed by a rigorous comparison of the different proposed wind patterns to the data. 

In the first section, we introduce the Multi-nested $\eta$\footnote{Forward modelling code from \cite{Ehrenreich2006}} Retrieval Code (MERC), highlighting the implementation of wind patterns, followed by a section dedicated to benchmarking the code on simulated data. We then move on to test the code on data obtained for HD~189733b and compare our findings to the current literature.

\section{Multi-nested ETA Retrieval Code (MERC)}
\label{sec:merc}
\begin{figure*}[htb!]
\resizebox{\textwidth}{!}{\includegraphics[trim=1.0cm 1.0cm 0.0cm 1.0cm]{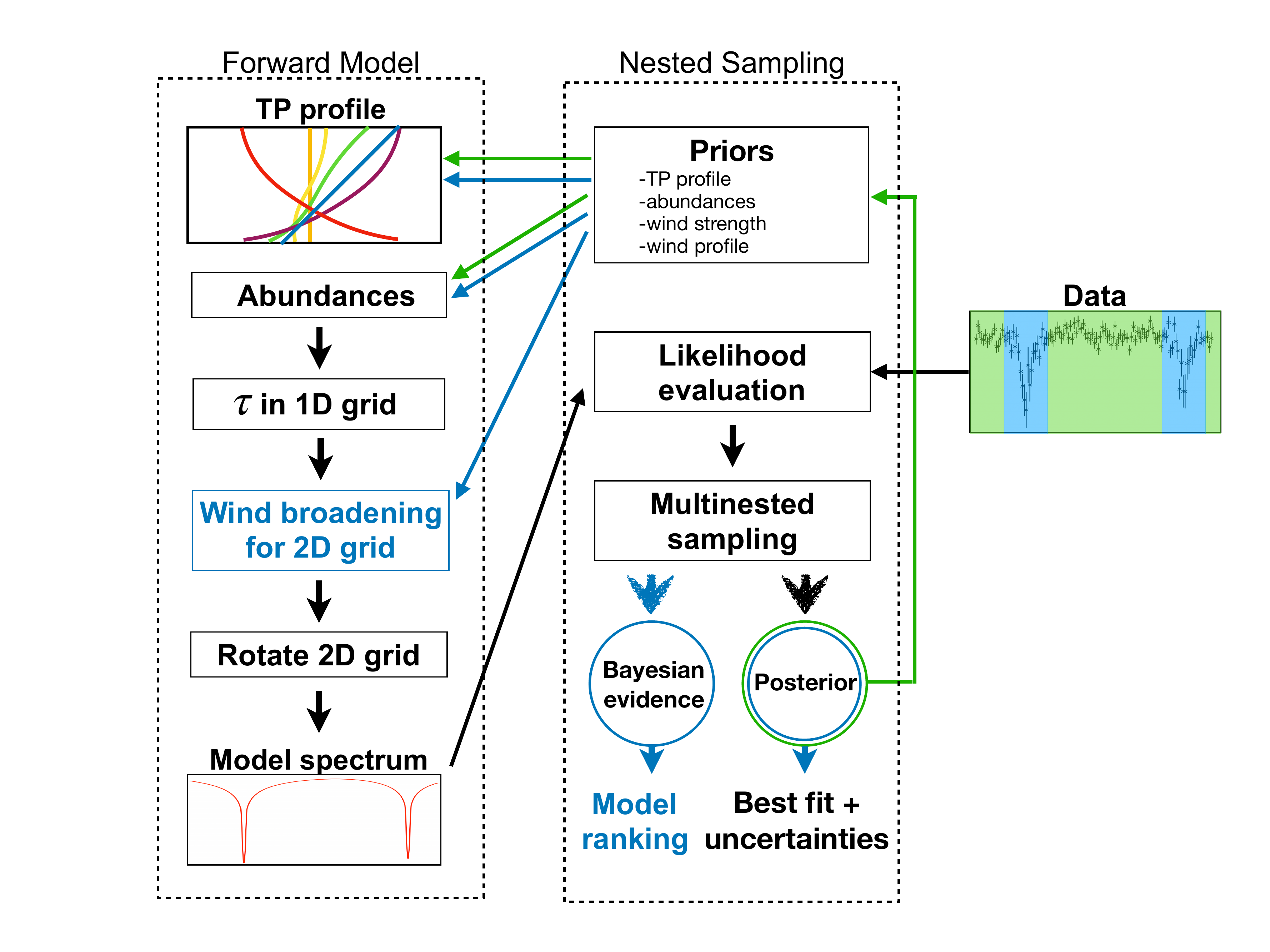}}
	\caption{Flowchart showing the information flow from forward model to the nested sampling module and vice versa in MERC. Green arrows are executed during the continuum retrieval, blue arrows are only executed during the line retrieval. Starting with the retrieval windows in green from Figure \ref{fig:retwindow} and the parameters that can be retrieved on the continuum (see Table \ref{table:MCMCoverview}), priors are set in the nested sampling module and handed to the forward model. One possible combination from the prior range is then used to calculate the extinction coefficient $\tau$ in a 1D grid in altitude above the planet surface. The planet surface is defined as the opaque part of the planet at the wavelength of the green region and is set at the optical white light radius. The 1D grid can be seen as a slice of the atmosphere, which is then used to create the full atmosphere (see Section \ref{sec:wind}) and calculate a model transmission spectrum. The model spectrum is handed back to the nested sampling module, where it is compared to the continuum of the data. The multi nested sampling executes this loop over the prior parameter space until the parameters converge to a value within the prior ranges. It then produces a posterior distribution of the prior and the Bayesian evidence of the model. The posterior of the continuum retrieval is then used to restrict the prior of the retrieval on the line retrieval window (shown in blue). The same process as for the continuum is started again, but now with additional wind broadening in each cell of the atmospheric slice. At the end the Bayesian evidence of the line retrieval for each model is used to rank them. This flowchart was loosely inspired on a similar layout in \cite{Brogi2019}.}
	\label{fig:flowchart}
\end{figure*}

 The structure of MERC is shown as a flowchart in Figure \ref{fig:flowchart} and consists of two parts, which are individually highlighted in this section: the forward model and the nested sampling. The forward model creates a model atmosphere based on input parameters and is an adaptation of the $\eta$ \citep{Ehrenreich2006} and subsequently the $^\pi\eta$ code \citep{Pino2018}, including wind broadening. This forward model is then linked to a nested sampling retrieval to explore the full parameter space. 

\subsection{Forward model}

In the forward model (left panel in Figure \ref{fig:flowchart}) a temperature pressure profile and element abundances are used as input to calculate a model transmission spectrum. Compared to the $\eta$ and $^{\pi}\eta$ code, the focus of the forward model in MERC lies on the execution speed. To combine a forward model with nested sampling retrieval, the forward model has to be executed more than $100~000$ times for each run, making a fast forward model indispensable. 

We achieve this speed-up by
\begin{itemize}
\item parallelizing and optimizing the code;
\item restricting the wavelength range to a single or few lines (in this case the sodium doublet);
\item calculating the extinction coefficient in 1D instead of 3D (already implemented in $^{\pi}\eta$).
\end{itemize}

 \begin{figure}[htb]
\resizebox{\columnwidth}{!}{\includegraphics[trim=-4.0cm 0.0cm 10.0cm 1.5cm]{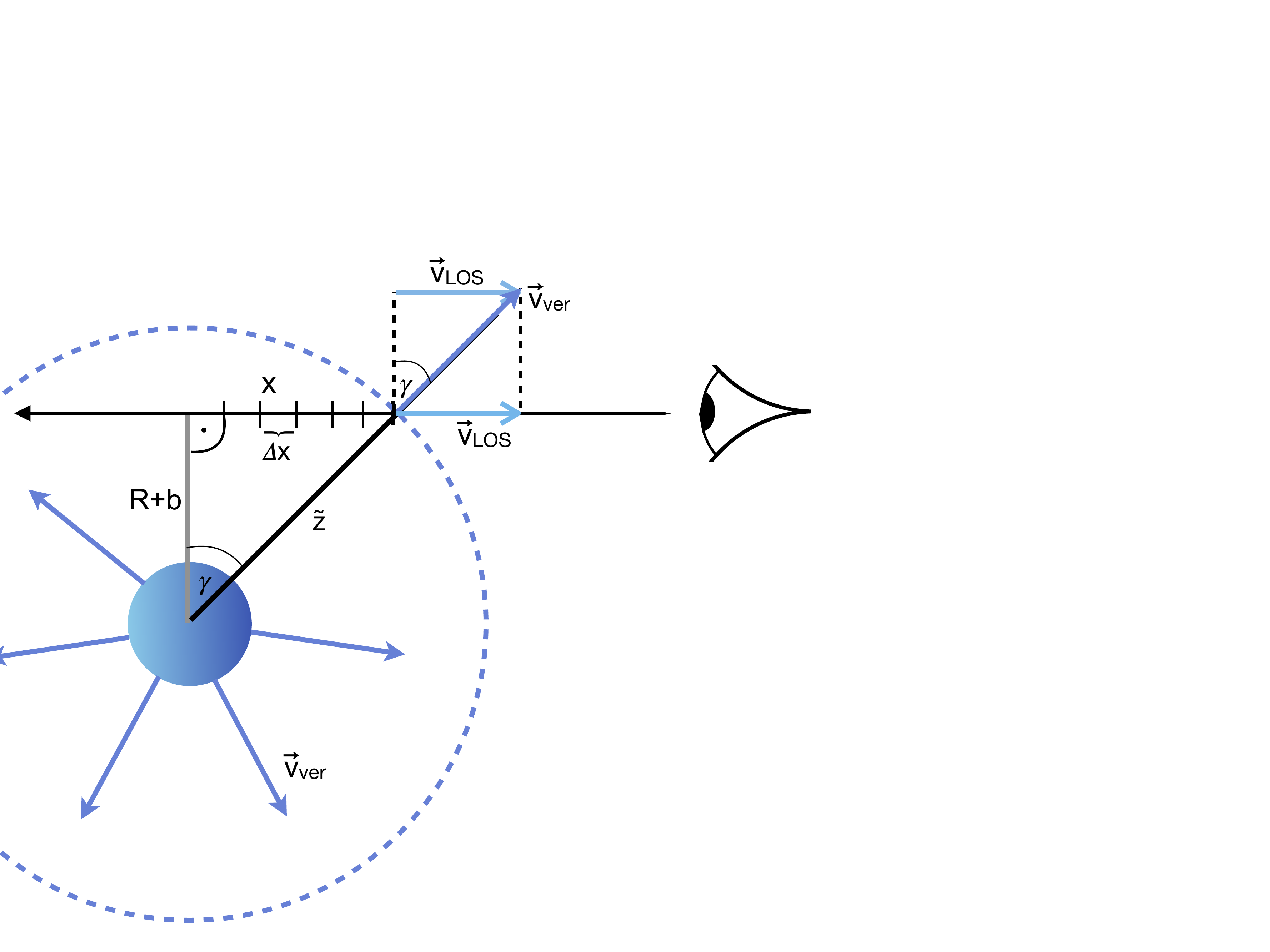}}
	\caption{Schematics of the extinction coefficient calculation in the line of sight (LOS) exemplified for the outermost bin in x for a vertical wind profile. The atmospheric wind pattern is assumed to be a constant radial velocity originating from the planet's surface. $\gamma$ is the angle running from the polar position towards the observer, $R$ the assumed radius of the planet, $b$ the impact factor and $x$ the sum over all bins in the line of sight $x=\sum \Delta x$}
	\label{fig:radvel}
\end{figure}
 
The atmosphere is described by a 1D grid in altitude above the planet surface. For computational reasons, an evenly spaced grid in log space with 200 cells covers the pressure from the assumed surface pressure (see Sec. \ref{sec:degen}) to $10^{-12}$ atm, far beyond the probing depth of the sodium doublet. The number of cells in pressure was set empirically as the lowest number of cells necessary to still have unchanged transmission spectra (difference $<1\%$) compared to runs with much finer grids. For each of these cells a temperature is calculated according to the different models. We explore an isothermal model and a temperature gradient under the assumption of hydrostatic equilibrium. Hydrodynamical models of hot Jupiter atmospheres show no better description of the data than hydrostatic models up to the lower part of the thermosphere, implying that the hydrostatic approach is a good assumption (e.g. \cite{Yelle2004,Koskinen2013}). Additionally, we are far below the sonic point of HD189733~b.
We implement the temperature gradient $T'$ linear in $\log(P)$ by setting a base temperature $T_{\mathrm{base}}$ at the base of our pressure scale and a top temperature $T_{\mathrm{top}}$ at the top of our pressure scale. 

 This temperature-pressure grid allows the calculation of the scale height, the gravity and the absolute height above the surface under the assumption of hydrostatic equilibrium. The estimation of the pressure at surface level is complicated by the degeneracy of the surface pressure and the sodium abundance, which we address in Section \ref{sec:degen}.
 
 From these environmental parameters, together with line parameters intrinsic to the sodium doublet -- the line center, Einstein coefficient $A_{21}$ and oscillator strength -- we calculate the cross section $\kappa_0$ of the sodium doublet in each cell of the atmospheric grid. The cross section grid is used to create an extinction coefficient grid of our atmosphere via:
 
 \begin{equation}
 \kappa(\lambda,z) = \chi(z) \cdot \kappa_0(\lambda,z)
 \label{eq:extinction_coefficient}
 \end{equation}
 
The extinction coefficient $\kappa$ is therefore the cross section of the molecule or element in question $\kappa_0$ weighted by the relative abundance of the element by mass, also known as the mass mixing ratio. In our case the element in question is sodium. The relative abundance $\chi$ is a parameter that we will also address in Section \ref{sec:degen}.
 
 At this stage, we calculated the extinction coefficient on a 1D grid corresponding to different altitudes. To compute the model transmission spectrum we need the slant optical depth ($\tau(\lambda,b)$) along the line of sight. The line of sight depends on the impact parameter $b(z)$ that quantifies how far from the planet surface the line of sight cuts through the atmosphere. To save significant computing time, the 1D grid of opacities in altitude is stored and for each calculation along the line of sight, the corresponding altitude from the surface ($\tilde{z}$ in Figure \ref{fig:radvel}) is calculated via the current impact parameter $b(z)$ and position along the line of sight (LOS) $x$.  $\tilde{z}$ is then used to select the corresponding extinction coefficient from the stored grid. This creates a two dimensional extinction coefficient grid $\kappa(\lambda,x,b)$ from the one dimensional $\kappa(\lambda,z)$, which we will call an atmospheric slice in the following. The geometry of this procedure is shown in Figure \ref{fig:radvel} and further elaborated on in \cite{Pino2018}. This two dimensional grid containing opacities in both z and x can then be modified by winds (see blue box in forward model in Figure \ref{fig:flowchart}), which we discuss in Section \ref{sec:wind}.
 
With the corresponding extinction coefficient in each bin, the slant optical depth is calculated by summing over all bins along the line of sight:
 
 \begin{equation}
\tau(\lambda,b) = \sum_{x} \kappa(\lambda,x,b)\cdot \Delta x
\end{equation}

The contribution of the atmosphere to the transmission spectrum, called the atmospheric equivalent surface of absorption ($\sum(\lambda)$), is created from the slant optical depths by iterating through the atmosphere with the line of sight in altitude ($z$). The altitude grid was defined earlier as a grid of $200$ cells from $z(i=0)=R_0$ to $z(i=200)$ being the top of the atmosphere linked to the pressure grid. The nomenclature for the atmospheric equivalent surface was selected to be in agreement with \cite{Ehrenreich2006}. 

The atmospheric equivalent surface of absorption from Eq. 3 of \cite{Ehrenreich2006} is:

\begin{equation}
\sum(\lambda) = \sum_{\mathrm{i=0}}^{200} 2\pi(b(i)+R_p){\tilde{z}(i)}(1-\exp[-\tau(\lambda,b(i))-\tau_{\mathrm{rayleigh}}(\lambda,i)])
\end{equation}

where the line of sight contributions are summed up for the full atmospheric contribution utilising the symmetry of the problem. $\tau_{\mathrm{rayleigh}}(\lambda,i)$ is the optical depth generated by Rayleigh scattering of $H_2$.

The transmission spectrum $\mathfrak{R}'(\lambda)$ is calculated as the contribution from the atmospheric equivalent surface and the opaque disk in LOS of the planet (Eq. $4$ from \cite{Ehrenreich2006}):

\begin{equation}
\mathfrak{R}'(\lambda) = - \frac{\sum(\lambda)+\pi R^2_P}{\pi R^2_*}
\end{equation}
where $\sum(\lambda)$ denotes the atmospheric equivalent surface of absorption and $\pi R^2_{P/*}$ respectively the opaque disk in LOS of the planet and the star.

In the next section we introduce an additional step when calculating the 2D atmospheric slice, a shift in the wavelength grid, which so far has remained constant. The wavelength grid is Doppler-shifted by wind patterns in the bins where the extinction coefficient is calculated, adding an overall broadening in the transmission spectrum.

 \subsubsection{Wind broadening}
\label{sec:wind}

\begin{figure*}[htb]
\resizebox{\textwidth}{!}{\includegraphics[trim=-2.0cm 2.0cm -2.0cm 2.5cm]{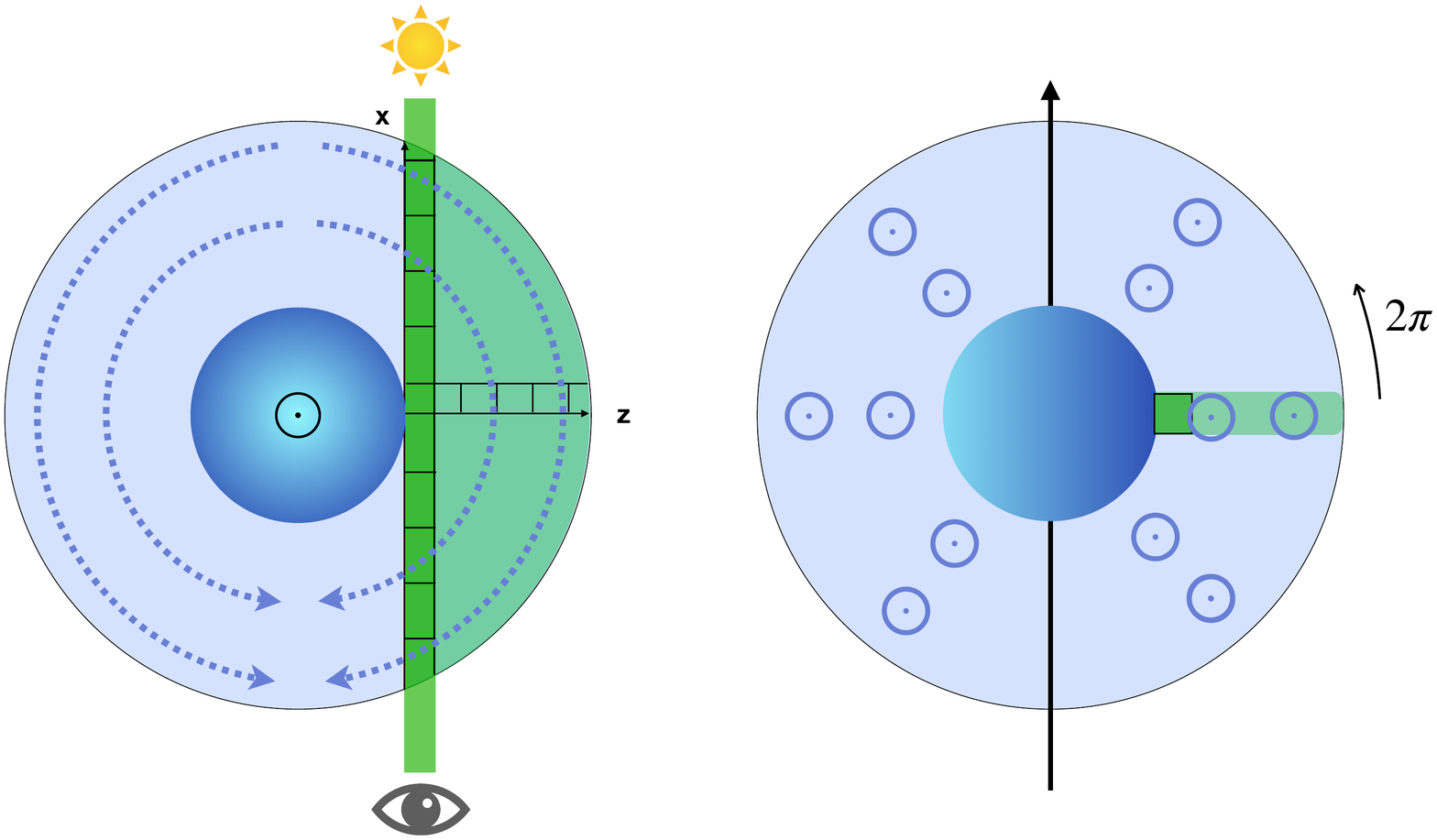}}
	\caption{Illustration of the implementation of the day-to-night side wind pattern with a 1D calculation of the atmosphere. A polar view is shown on the left and an equatorial view on the right. The wind is indicated by the dotted blue arrows. \textbf{Left:} The extinction coefficient is calculated in altitude along the z axis and then transposed in the x direction along the LOS (in dark green). The LOS is then iterated upward in z until the top of the atmosphere is reached and all values saved in a 2D grid (here visualised as a slice in light green). In each bin of the 2D grid the velocity and the broadened profile are calculated and stored. This is only necessary on one side of the atmosphere due to the symmetry of the problem. \textbf{Right:} The calculated slice is then rotated to create the full atmosphere. The line of sight is shown as a dark green box in the main illustration, where the reader is in the position of the observer. Points indicate a flow towards the reader. In the day-to-night side case, the morning and evening limb winds point towards the observer and only one atmospheric slice has to be calculated. The slice is then rotated by $2\pi$ to create the full atmosphere. In this simplification of the atmosphere the wind will not go parallel to the equator at all times, but point towards the center of the night side. This reduces calculation time significantly, given that it reduces the problem from 3D to 2D, with the extinction coefficient only calculated in 1D.}
	\label{fig:dtnillustration}
\end{figure*}

\begin{figure*}[htb]
\resizebox{\textwidth}{!}{\includegraphics[trim=-2.0cm 2.0cm -2.0cm 2.0cm]{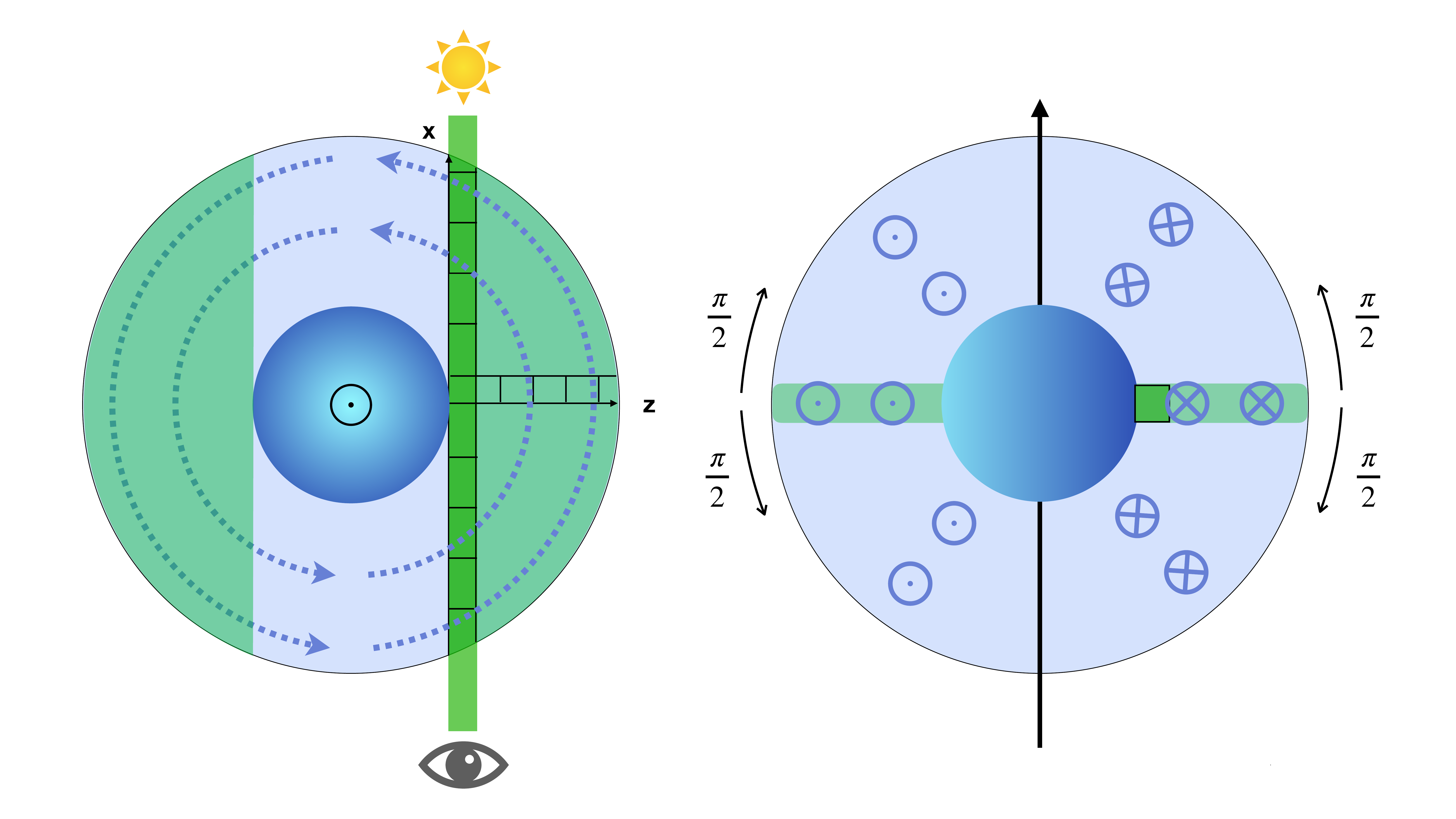}}
	\caption{Illustration of the implementation of the super-rotational pattern with a 1D calculation of the atmosphere. See Figure \ref{fig:dtnillustration} for a detailed description of the figure and geometrical consequences on the computation. \textbf{Left:} Since the wind is coming towards the observer for the morning terminator and away form the observer for the evening terminator, a slice of the atmosphere has to be calculated for each.  \textbf{Right:} The calculated slices are then rotated by $\pi$ to get the full atmosphere (one half each). In the current model the wind does not change with latitude, creating a super-rotational stream that is constant tangentially over each atmospheric shell.}
	\label{fig:srotillustration}
\end{figure*}

\begin{figure*}[htb]
\resizebox{\textwidth}{!}{\includegraphics[trim=-2.0cm 2.0cm -2.0cm 2.5cm]{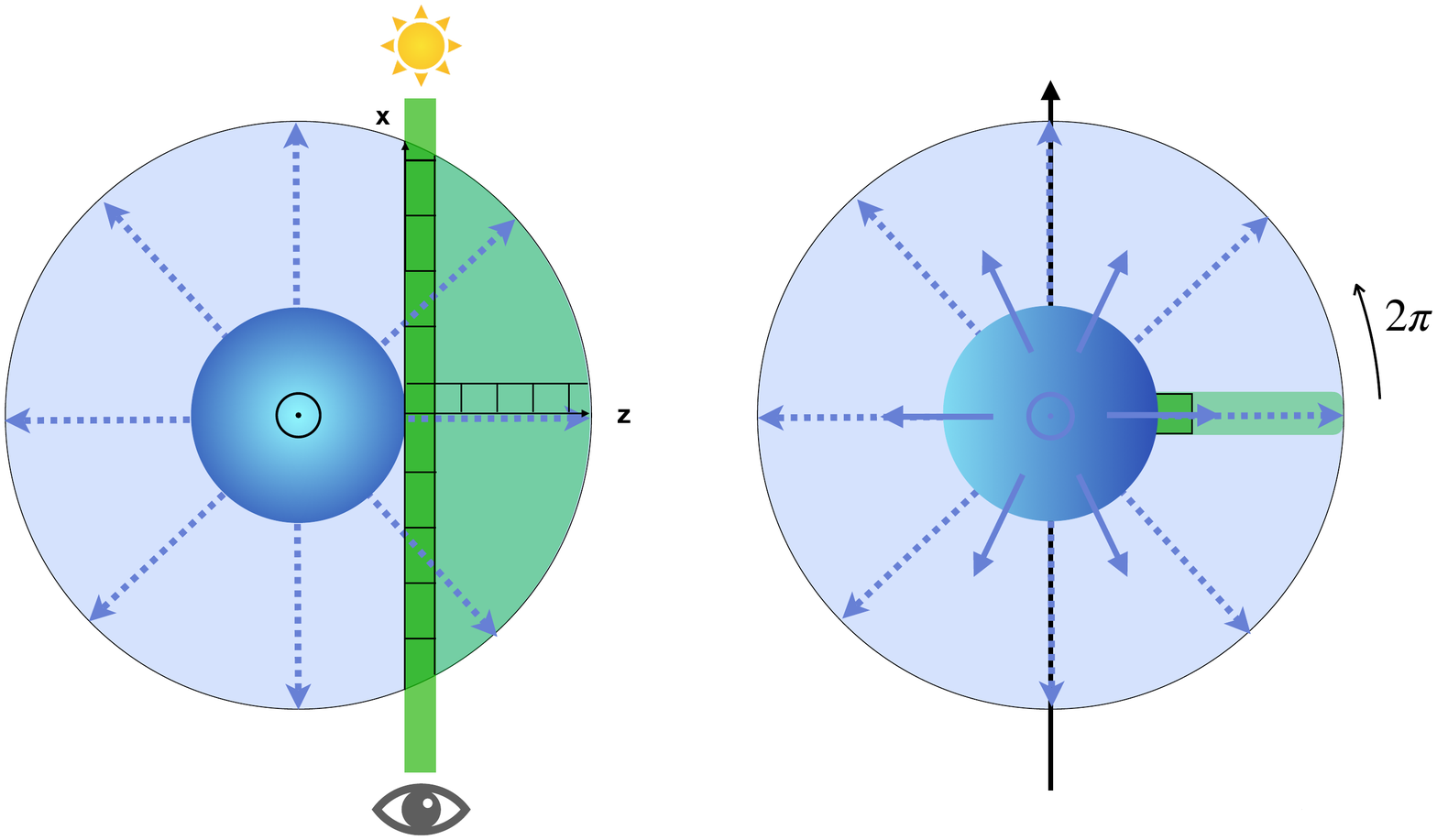}}
	\caption{Illustration of the implementation of the vertical pattern with a 1D calculation of the atmosphere. See Figure \ref{fig:dtnillustration} for a detailed description of the figure and geometrical consequences on the computation. \textbf{Left:} The calculation of the LOS extinction coefficients is only necessary on one side of the atmosphere due to the symmetry of the problem. \textbf{Right:} The calculated slice is then rotated by $2\pi$ to get the full atmosphere. The constant wind points outwards for all latitudes.}
	\label{fig:verillustration}
\end{figure*}

When traversing the atmosphere towards the observer the wavelength of the light is Doppler shifted using the amplitude of the velocity in the line of sight $\vec{v}_{\mathrm{LOS}}$ in each bin in x. This velocity stems from winds in the atmosphere, since the transmission spectrum is calculated in the planet rest frame. To account for these winds only, a correction for the different velocities originating from the different rest frames of observer and observed object is required. The three wind patterns most interesting for hot Jupiters are a day-to-night side (dtn) wind (see Figure \ref{fig:dtnillustration}), a super-rotational wind (see Figure \ref{fig:srotillustration}) and a vertical upward wind (see Figure \ref{fig:verillustration}), which we study separately and in combination in the following. 

First, we demonstrate how the velocity shift is calculated in a specific bin for the three different wind patterns. 
We exemplify the full calculation for a simple upward pointing constant wind velocity, see also Figure \ref{fig:radvel} for a visualization.

Assuming a constant vertical wind originating from the planet atmosphere the line of sight velocity depends on the constant amplitude of the wind ($|\vec{v}_{\mathrm{ver}}|$) and on the angle of the wind vector and the line of sight $\vec{v}_{\mathrm{LOS}}=|\vec{v}_{\mathrm{ver}}|\cdot \sin \gamma $

With $\tilde{z}=\sqrt{(R+b)^2+x^2}$, the distance of the current bin from the planet center (see Figure \ref{fig:radvel}), the angle $\gamma$ is calculated from the similarity of the two triangles in Figure \ref{fig:radvel} as
\begin{equation}
\sin \gamma = \frac{x}{\tilde{z}}=\frac{x}{\sqrt{(R+b)^2+x^2}}
\end{equation}

leading to a velocity in line of sight depending only on the x and y position on the grid, which will be iterated through in the code.
For a vertical constant wind the wind strength in the line of sight is calculated as follows:

\begin{equation}
|\vec{v}_{\mathrm{LOS}}|=|\vec{v}_{\mathrm{ver}}|\cdot \frac{x}{\sqrt{(R+b)^2+x^2}}
\end{equation}

A super-rotational wind pattern throughout the atmosphere and a day-to-night side wind are calculated the same way and the sign is adjusted depending on the respective wind direction in each hemisphere:

\begin{equation}
|\vec{v}_{\mathrm{LOS}}|=\pm|\vec{v}_{\mathrm{rot/dtn}}|\cdot \frac{(R+b)}{\sqrt{(R+b)^2+x^2}}
\end{equation}

The Doppler shift of the wavelength grid is calculated for each bin by
\begin{equation}
\lambda_{\mathrm{shifted}}=\lambda_{\mathrm{grid}}\cdot \left(1+\frac{|\vec{v}_{\mathrm{LOS}}|}{c}\right)
\end{equation}
where $c$ is the speed of light. 

Depending on which kind of velocity profile is implemented, the line of sight velocity changes for each bin and via its Doppler broadening the entire transmission spectrum. The three Figures \ref{fig:dtnillustration}, \ref{fig:srotillustration} and \ref{fig:verillustration} show how the velocity shift in the line of sight (dark green bins, left side of each figure) is translated into a pseudo-3D wind pattern (right side of each figure). 

In the specific case of a radial velocity field, the symmetry of the field can be utilized by calculating the shift for the side towards the observer with $+|\vec{v}_{\mathrm{LOS}}|$ and the side facing away from the observer with $-|\vec{v}_{\mathrm{LOS}}|$ (see Figure \ref{fig:verillustration}). The cases of a super-rotational wind and a day-to-night side wind are less intuitive. For the day-to-night side wind (see Figure \ref{fig:dtnillustration}) both the southern and the northern hemisphere, and the morning and evening limb, behave in the same way and we rotate our line of sight calculation to compute the entire atmosphere. This implies the generalisation of a constant day-to-night side wind throughout the atmosphere without the possibility for stronger winds at the equator and weaker winds at the poles. As a consequence, the day-to-night side wind points towards the center of the night side and does not go parallel to the equator. The same simplification applies to super-rotational winds (see Figure \ref{fig:srotillustration}). This is a direct consequence of our computational approach, which speeds up the execution of the code by orders of magnitude. However, we would like to highlight that the unrealistic approach of constant winds means in the case of a super-rotational wind that the actual wind speed within the jet is likely higher than our retrieved value.

Yet, more realistic models of atmospheres rarely show only one dominant wind pattern. We will therefore study combined wind patterns, where a wind pattern is assumed in the lower atmosphere (either super-rotation of day-to-night side) and another one in the upper atmosphere (in this case vertical upward winds). We express the switch in orders of magnitude in altitude above the surface of the planet:

\begin{equation}
\label{eq:pswitch}
P_{\mathrm{switch}} = P_0\cdot10^{h}
\end{equation}
where $h$ is the variable for the retrieval ($h<0$) and $P_0$ is the surface pressure deduced from NaX (see Section \ref{sec:degen}).

As a consequence of the Doppler shift of the profiles, a different wavelength grid is created in each bin. For the subsequent summation over all bins to produce the transmission spectrum, we have to interpolate the profile in each bin on a reference wavelength grid. We choose a reference grid at the same resolution as the HARPS wavelength calibration.
The simplification of the wind profiles allows an extinction coefficient calculation in 1D, while retaining most of the 3D information of the wind patterns. This in turn increases computational efficiency to a degree where the forward model, including the broadening from winds, can be combined with a nested sampling retrieval.

\subsection{Nested sampling retrieval}
\label{sec:retrieval}

Once the forward model creates a transmission spectrum for the sodium doublet, we use a nested sampling retrieval to find the best-fitting model from the selection. 

We couple a nested sampling algorithm to the described forward model to explore the full parameter space of all variables (see Table \ref{table:MCMCoverview}). The main advantage of nested sampling lies in the direct calculation of the Bayesian evidence. This calculation, taking into account different parametrisations with changing number of parameters, allows to compare these models equivalently \citep{Skilling2006,Feroz2009,Benneke2013,Waldmann2015,Line2016}, in a direct application of Occam's Razor. The basics of these concepts are introduced in \cite{Trotta2008, Skilling2006}. A full description of the approach taken here can be found in \cite{Lavie2016}.  For similar applications and examples see \cite{Benneke2013,Waldmann2015,Line2016}. 

To begin, $N_{\mathrm{live}}$ points are randomly selected from the parameter space under the constrain of our prior, where N parameters form an N-dimensional parameter space. For this first set of points, the likelihood values are calculated and in each step, the algorithm discards the point with the lowest likelihood and adds a new one until convergence is reached (see \cite{Skilling2006,Lavie2016}). Taking into account the independent Gaussian errors of our spectral observations, we calculate the likelihood as a Gaussian function (Equation 3 in \citep{Benneke2013}. As in the implementation of \cite{Lavie2016}, we use the open-source software {\tt PyMultiNest}\footnote{https://github.com/JohannesBuchner/PyMultiNest/}\citep{Buchner2016}, a wrapper for the open source {\tt MultiNest}\footnote{https://ccpforge.cse.rl.ac.uk/gf/project/multinest/} module \citep{Feroz2008,Feroz2009,Feroz2013}. 

Typical numbers of $N_{\mathrm{live}}$ range from $N_{\mathrm{live}}=50$ to $10,000$ \citep{Benneke2013} up to $40,000$ (accumulated, 400 runs with 100 live points) in \cite{Lavie2016}, in this paper we will use $10,000$ (accumulated, 100 runs with 100 live points) each. 

The Bayesian evidence for each of these models is calculated via Eq. 4 in  \cite{Lavie2016} and allows us to compare the different models (here model $M_0$ and $M_1$ with Bayesian evidence $Z_0$ and $Z_1$) to each other via the logarithm of their Bayes factor \citep{Trotta2008}:

\begin{equation}
\label{eq:Bayes}
\ln \mathcal{B}_{01}= \log (Z_0/Z_1)= \log Z_0 - \log Z_1
\end{equation}

The Bayes factor is then judged via the so called 'Jeffrey's scale', an empirical scale shown in Table \ref{table:Bayesianoverview} \citep{Trotta2008}. It highlights when $\ln \mathcal{B}_{01}$ presents weak, moderate or strong evidence in favour of $M_0$ over $M_1$. The best-fit model allows us to rule out certain atmospheric scenarios, while taking into account the complexity of the different models and thus make quantitative statements about winds and temperature profiles in specific exoplanet atmospheres.

\begin{table}
\caption{Empirical scale to judge the evidence when comparing two models $M_0$ and $M_1$, called a 'Jeffrey's scale'. The following table together with a more in depth explanation can also be found in \cite{Lavie2016}.}
\label{table:Bayesianoverview}
\centering
\begin{tabular}{l l l l }
\hline
\hline
$|\ln\mathcal{B}_{01}|$   & Odds & Probability & Strength of evidence     \\
\hline
$<1.0$  &  $<3:1$   &  $<0.750$  &   Inconclusive \\
$1.0$  &  $\sim 3:1$   &  $0.750$  &  Weak evidence \\
$2.5$  &  $\sim 12:1$   &  $0.923$  &   Moderate evidence \\
$5.0$  &  $\sim 150:1$   &  $0.993$  &   Strong evidence \\
   \hline
\end{tabular}
\end{table}

\subsection{Combination in MERC}
\begin{figure}[htb!]
\resizebox{\columnwidth}{!}{\includegraphics[trim=0cm 0.5cm 2.0cm 0.5cm]{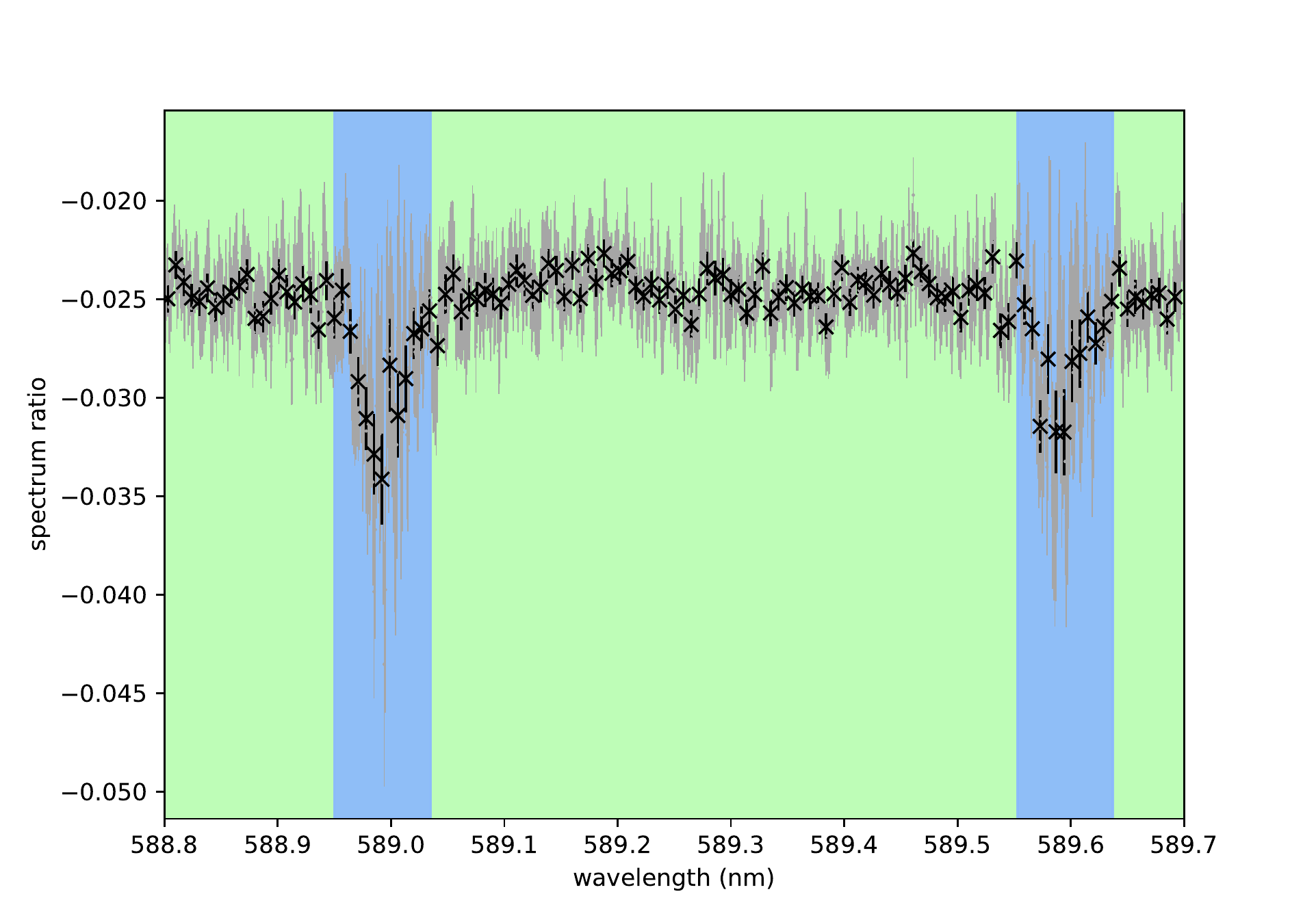}}
	\caption{Spectral ratio of the hot Jupiter HD189733b for the sodium doublet. In grey the data obtained with the HARPS spectrograph, in black the binned data for visibility (for a detailed description of the data see Section \ref{sec:data}). The dark, blue areas highlight the retrieval zones at line center and the light, green areas the retrieval on the continuum.}
	\label{fig:retwindow}
\end{figure}
In MERC, we combine the forward model with the multi nested sampling module by using the priors of the nested sampling to create the temperature pressure (TP) profile and abundances needed as input for the forward model. The code then takes the result of the forward model, the model transmission spectrum, and hands it back to the nested sampling where the Bayesian evidence of the model is computed.

For the Bayesian evidence, MERC analyses the difference between the value of the model and the value of the data in each wavelength bin, meaning that each bin carries the same weight for all parameters. This comparison is performed on the unbinned HARPS data to preserve accuracy.
Yet, the temperature and wind strength have only a strong impact on the line core and the visible line wings, a small subset of the total number of bins in the sample, whereas the pressure at the base of the atmosphere and the constant sodium abundance are driven by the data continuum with a negligible impact from the core.
To address this issue we introduced a two tier approach to the standard nested sampling retrieval, where we first retrieve the parameters dominated by the location of the continuum and then retrieve the additional parameters dominated by the line core.

In the first step, the data points of the sodium doublet core are blinded, by leaving out the wavelength range twice the full width half maximum (FWHM), which includes the line cores. In the second step only this area containing the doublet is used for the retrieval. The two areas are highlighted in Figure \ref{fig:retwindow}, where blue shows the line core retrieval range and green the continuum retrieval range. From here on ,we call these two steps continuum retrieval and line retrieval. The retrieval windows are chosen on an empirical basis, to include all data points from the line core until the inclusion of additional points does not significantly change the retrieval results. We benchmark this approach by running the retrieval on line retrieval windows of 3 and 4 FWHM of the lines, with no changes to the result, but a significant increase in runtime. 
 
 In the flowchart of MERC (Figure \ref{fig:flowchart}), the split of the data into continuum and line core is highlighted with the first, green cycle of the algorithm, the continuum retrieval, followed by the blue cycle, the line retrieval. The posterior of the continuum retrieval is used to set the prior for the line retrieval and only the line retrieval is affected by winds. The Bayesian evidence of the line retrieval is then used to rank the models and the posterior of the line retrieval gives the best-fit parameters and their uncertainties. The uncertainties are calculated as $1\sigma$ of the distribution. Given that not all posteriors follow a normal distribution, the two limits are calculated via the $0.16$ and $0.84$ quantile to include $68\%$ of the distribution. The best-fit is calculated as the bin in the posterior distribution that had overall the highest Bayesian evidence. All corner plots for the posteriors with their indicated best-fits can be found in the Appendix \ref{app:HD189}.
 
MERC is particularly well suited for the wavelength range of the sodium doublet, given its intrinsic large absorption depth, thus probing layers up to the thermosphere of the exoplanet. The different models employed to describe the atmosphere up to the thermosphere are listed in Table \ref{table:MCMCoverview}, where all retrieved parameters are shown. The continuum parameters are solely used to restrict the prior of the line retrieval and are still retrieved on the line.

\begin{table}
\caption{Overview of all models and retrieved parameters. The temperature base models can be combined with one or more of the additional wind patterns. All continuum parameters are also retrieved on the line, with their prior restricted by the continuum posterior. \NaX is the degenerate parameter encompassing the pressure and sodium abundance and is defined in \ref{eq:nax}.}
\label{table:MCMCoverview}
\centering
\begin{tabular}{l c c}
\hline
\hline
Model      &  Line & Continuum        \\
\hline
isothermal & \NaX, $T_0$ & \NaX, $T_0$\\
T gradient & \NaX, $T_0, T_1$ & \NaX, $T_0$ \\
\hline
vertical wind & \NaX, $T_0$,$v_{\mathrm{vertical}}$ & \NaX, $T_0$ \\
super-rotation & \NaX, $T_0$,$v_{\mathrm{rot}}$ & \NaX, $T_0$ \\
day-to-night side & \NaX, $T_0$, $v_{\mathrm{dtn}}$ &\NaX, $T_0$ \\
   \hline
combined wind patterns & \NaX, $T_0$, $v_{\mathrm{low}}$, $v_{\mathrm{up}}$, $h$ & \NaX, $T_0$ \\
   \hline
\end{tabular}
\end{table}

\subsection{Impact of the abundance pressure degeneracy}
\label{sec:degen}

Based on the transmission spectrum geometry of \cite{Ehrenreich2006} and the theoretical work of \cite{Fortney2005}, \cite{Lecavelier2008} derived the effective altitude depending on wavelength as a function of extinction coefficient and partial pressure of the main absorber. Taking into account the variation of the cross section as a function of wavelength, a relation we also make use of in MERC to create a model transmission spectrum from atmospheric input parameters, they were able to relate the abundance of the absorber species to the temperature grid in altitude. However, this relation depends on the effective planetary radius at z=0, $R_0$, and subsequently on the pressure at z=0, $P_0$. Additionally,  \cite{Heng2017} inferred a three way degeneracy between $P_0$, $R_0$ and the element abundances $\chi$, all of which change the level of the continuum in the transmission spectrum.

In more recent work, \cite{Fisher2018}  showed, that this three way degeneracy is partially due to model assumptions in semi-analytical approaches, and \cite{Welbanks2019} that the degeneracy with the element abundances can be broken in retrieval analysis. This would require a more rigorous treatment of parameters influencing the continuum of the transmission spectra than presented here, most importantly the inclusion of $H_2-H_2$ collision induced absorption (CIA), but also cloud decks, hazes and additional absorber species. 

We follow \cite{Lecavelier2008} in their description on the relation between the pressure scale and the height scale, assuming hydrodynamical equilibrium, and set the reference radius at $P_0$ to the white light radius. We then retrieve $P_0$ to break the degeneracy between $P_0$ and $R_0$. \cite{Welbanks2019} verified this approach and showed that $R_0$ can be set to any measured value, because the retrieval will adjust $P_0$ accordingly.

Taking into account the simplicity of our forward model and the new findings by \cite{Welbanks2019}, we combine the two degenerate parameters $P_0$ and $\chi$ to set the continuum:

\begin{equation}
\NaX = \frac{\chi}{\chi_{\astrosun}} \cdot \frac{P_0}{10~\mathrm{bar}} = 10^{-5+\mathrm{NaX}}
\label{eq:nax}
\end{equation}

The right-hand side in Equation \ref{eq:nax} shows how the degenerate parameter is retrieved in MERC, where only the exponent that indicates the change from default values is a parameter. If NaX is $0$, $\NaX$ is set to the default values.
The abundance is then fixed throughout the atmosphere to a constant value, which is a commonly adopted approximation \citep{Lecavelier2008,Agundez2014,Steinrueck2019}. The shape of the continuum is modelled via the calculation of the Rayleigh spectrum for $H_2$ \citep{Lecavelier2008}.

We consider the presented simplistic approach following \cite{Heng2015} to treat the continuum sufficient for the purpose of this work, where the focus lies on the line shape and its possible origin in atmospheric wind patterns. The line shape is not significantly impacted by any of the degenerate parameters of this section. However, in follow up studies we plan to implement a more realistic approach in describing the element abundances, and the relation between pressure and height in atmospheric retrievals, following \cite{Welbanks2019}, to show their impact on line broadening.

\section{Test on simulated data}

We test MERC (see Section \ref{sec:merc}) on simulated transmission spectroscopy data. We constructed two different simulated transmission spectra for HD~189733b with the HARPS resolution and the mean divergence and error of a HARPS transmission spectrum. We use the mean spread of the data in the continuum to introduce noise to the simulated data. MERC then creates one transmission spectra for HD~189733b with a purely isothermal atmosphere at $3600$~K and no winds, and a second transmission spectra with an isothermal temperature profile at $3600$~K and an added vertical wind throughout the atmosphere with a wind speed of $30~\kms$. In both datasets, NaX was set to $-1.3$ to set the continuum. Both simulated datasets were created for arbitrary values of the parameters, one without and one with wind to highlight the ability of MERC to distinguish between a base model and a model with additional parameters. We then ran MERC on both simulated data sets retrieving on each a isothermal temperature profile with no winds, and an isothermal profile with an added vertical wind and compared if the code favoured the correct model for the simulated data sets.

\subsection{Isothermal atmosphere}

An overview of the parameters used to create the simulated data without winds and the two models is shown in Table \ref{table:simiso}. The isothermal model is a better fit than the model with vertical wind, but the difference in Bayesian evidence is inconclusive. Both models provide adequate fits to the simulated data since the model with added vertical winds centres the added wind around $0$, making them equal (see Figure \ref{fig:simisoplot}). The more complex model with the added wind is then punished for the additional retrieval parameter and has therefore a lower Bayesian evidence. Both parameters T and NaX are retrieved correctly as their original values.

\begin{table*}
\caption{Comparison of the different models for a simulated transmission spectrum from an isothermal atmosphere.}
\label{table:simiso}
\centering
\begin{tabular}{l c c c c c}
\hline
\hline
 & $|\ln\mathcal{Z}|$ &   & T [K] & NaX & $|\vec{v}_{\mathrm{ver}}| [\kms]$    \\
\hline
sim. data  &  &   & $3600~K$  & $-1.3$  &  $0$ \\
\hline
isothermal model  &  $793.94\pm0.16$& mean$\pm1\sigma $& $3459^{+314}_{-376}$  & $-1.21^{+0.36}_{-0.25}$  &  -\\
  & & best-fit & $3599$  & $-1.31$  &  -\\
  \hline
   + vertical wind  &  $792.93\pm0.18$& mean$\pm1\sigma $& $3314^{+365}_{-449}$  & $-1.10^{+0.44}_{-0.30}$  &  $3.34^{+7.36}_{-6.19}$\\
  & & best-fit &  $3545$ & $-1.27$  & $-2.11$ \\
   \hline
\end{tabular}
\end{table*}

\subsection{Isothermal atmosphere with vertical winds}

In the second test, we create a simulated dataset from a model spectrum that includes vertical upward winds at $30~\kms$. We then employ MERC to model the dataset with an isothermal model and an isothermal model with additional winds, the same model used to create the data. The parameters with the posteriors from both retrievals are presented in Table \ref{table:simwind}. It shows that the isothermal model by itself is unable to sufficiently fit the provided simulated data with a Bayesian evidence $5.5$ lower than the evidence for the isothermal model with added winds. This Bayesian evidence is strong evidence to support the second model (see Table \ref{table:Bayesianoverview}). The second model with the added wind is additionally also able to retrieve all parameters to the correct values (best-fit compared to sim data in Table \ref{table:simwind}). 

\begin{table*}
\caption{Comparison of the different models for a simulated transmission spectrum from an isothermal atmosphere with vertical upward wind.}
\label{table:simwind}
\centering
\begin{tabular}{l c c c c c}
\hline
\hline
 & $|\ln\mathcal{Z}|$ &   & T [K] & NaX & $|\vec{v}_{\mathrm{ver}}| [\kms]$    \\
\hline
sim. data  &  &   & $3600~K$  & $-1.3$  &  $30$ \\
\hline
isothermal model  &  $776.54\pm0.17$& mean$\pm1\sigma $& $3674^{+217}_{-323}$  & $-1.05^{+0.30}_{-0.19}$  &  -\\
  & & best-fit & $3884$  & $-1.22$  &  -\\
  \hline
   + vertical wind  &  $791.94\pm0.19$& mean$\pm1\sigma $& $3554^{+259}_{-303}$  & $-1.29^{+0.31}_{-0.23}$  &  $30.33^{+3.42}_{-3.25}$\\
  & & best-fit &  $3633$ & $-1.38$  & $30.2$ \\
   \hline
\end{tabular}
\end{table*}

MERC is able to correctly model both simulated datasets and rank different models according to their ability to model the data. The posteriors of the retrieved parameters and the best-fit models together with the simulated data are shown in the Appendix in Figure \ref{fig:simposterior} and Figures \ref{fig:simisoplot} and \ref{fig:simverplot}.

\section{Benchmarking with HD~189733b}

\subsection{Data reduction}
\label{sec:data}
We apply MERC to HD~189733b, one of the most well studied exoplanets to date. 
We use data from the HARPS echelle spectrograph mounted at ESO's 3.6m telescope in la Silla, Chile. The calculation of the transmission spectrum and the telluric correction were undertaken by \cite{Wyttenbach2015}. However, they did not correct the data for the Rossiter-McLaughlin (RM) effect. The RM effect induces  distortions in the transmission spectrum due to the rotation of the star (e.g. see \cite{Rossiter1924,McLaughlin1924,Cegla2016,Tr18}). 

\begin{figure}[htb!]
\resizebox{\columnwidth}{!}{\includegraphics[trim=3.0cm 9.0cm 3.0cm 9.0cm]{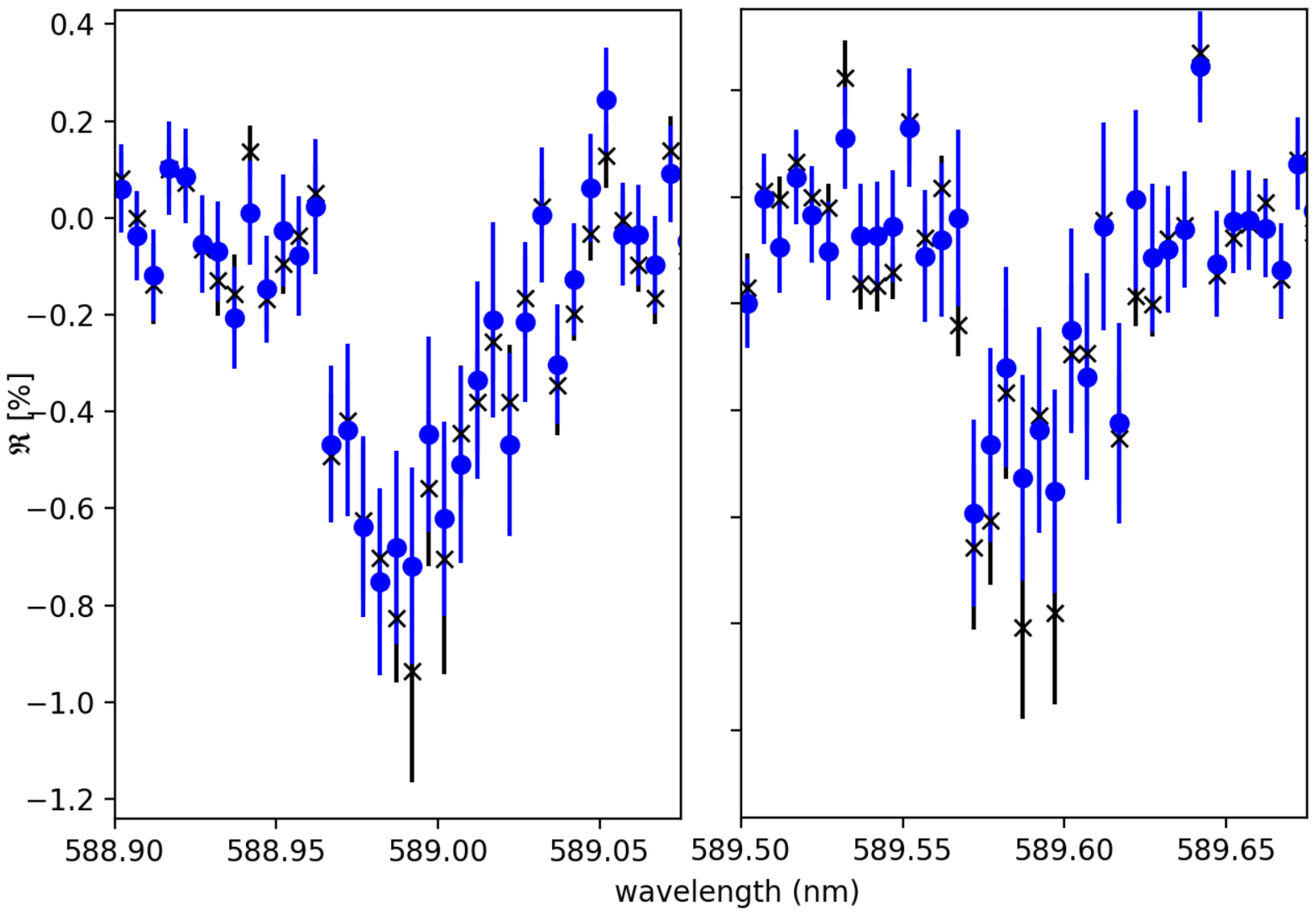}}
	\caption{Transmission spectrum of the two sodium doublet lines binned by x5. Both spectra were generated from HARPS data of HD~189733b. In black is the spectrum from \cite{Wyttenbach2015} with the additional correction of the RM effect. In dark blue is the RM corrected spectrum from \cite{Casasayas-Barris2017} which we use as a reference with permission. The offset between the two high-resolution transmission spectra is consistently below $1\sigma$ and the smallest in the line wings, which is the most important range for wind retrieval. We therefore consider our RM correction to be sufficient for our purposes.}
	\label{fig:RM}
\end{figure}

We removed the stellar RM signature by dividing with the sum of all out-of-transit spectra (master out) after shifting it to the local velocities of the star  behind the  planet. The local stellar velocity was derived from \cite{Cegla2016}. We compared the result to the RM corrected sodium transmission spectrum for HD~189733b from \cite{Casasayas-Barris2017}. The two datasets were overplotted (see Figure \ref{fig:RM}) and points are within $1\sigma$ of the planet transmission spectrum of the other dataset, showing that our RM correction is sufficiently accurate. 

\subsection{Retrieval results}
\label{sec:resultsHD189}
We used MERC on multiple models: an isothermal temperature profile, a temperature gradient, an isothermal profile combined with different wind patterns: a day-to-night side wind, a super-rotational wind, a vertical upward wind and a combination of lower atmosphere day-to-night or super-rotational wind with vertical upward wind in the upper atmosphere.

The Bayesian evidence from all the models can be found in Table \ref{table:HD189733bcomparison}. All posterior distributions are in Appendix \ref{app:HD189}, with the best-fit marked in blue and the mean values of the posterior printed on top of each column. 

The isothermal model, as the simplest model, is used as the base model for comparison and has a Bayesian evidence of $|\ln\mathcal{Z}|=726.92\pm0.14$. A comparison of all different models to each other is shown in Figure \ref{fig:retrievalresults}, where the upper row of models is compared to all other models. Red stands for no evidence for a better fit, yellow for moderate evidence and green for strong evidence. The comparison scheme can be found in Table \ref{table:Bayesianoverview}. 

\subsubsection{Isothermal profile versus temperature gradient}

\begin{figure}[htb!]
\resizebox{\columnwidth}{!}{\includegraphics[trim=3.0cm 8.0cm 3.0cm 9.0cm]{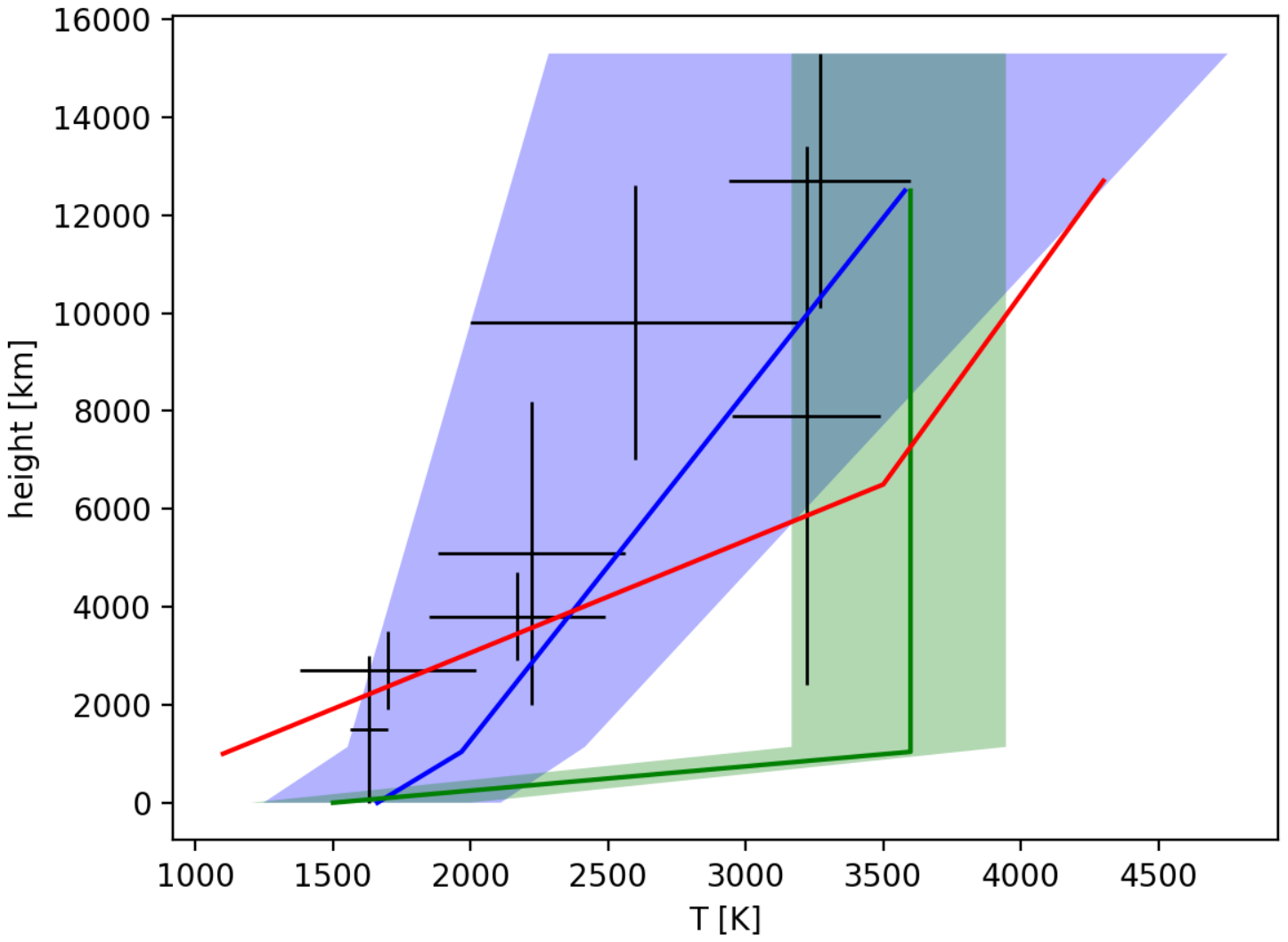}}
	\caption{Temperature pressure profiles for HD~189733b. The points with error bars in black are from \cite{Wyttenbach2015}, the red line corresponds to the profile from \cite{Huang2017} and the green and blue lines are the isothermal (see Figures \ref{fig:isobest} and \ref{fig:isoposterior}) and temperature gradient models (see Figures \ref{fig:gradbest} and \ref{fig:gradientposterior2}) from this work. The shaded area corresponds to the $1\sigma$ area of the posterior distribution. The lowest data point in height comes from the continuum retrieval, the rest of the curve from the line cores.}
	\label{fig:TPprofiles}
\end{figure}

Comparing the isothermal and temperature gradient models shows that there is no evidence for a better fit with the more complex temperature gradient. The posterior distributions for both models are in the appendix, Figures \ref{fig:isoposterior} and \ref{fig:gradientposterior}. The mean of the isothermal temperature is $T(\mathrm{mean}) = 3412^{+347}_{-432}$~K. The posterior of the temperature gradient shows a wide spread for the base temperature at the surface of the planet $T_{\mathrm{base}}(\mathrm{mean})= 2191^{+606}_{-858}$~K and a best-fit close to the edge of the prior ($T_{\mathrm{base}}= 3182$~K). The temperature at the top of the atmosphere shows a similarly wide spread. The comparatively poor convergence for this model stems from the degeneracy between the continuum parameter $\NaX$, the base temperature $T_{\mathrm{base}}$, and subsequently the top temperature $T_{\mathrm{top}}$. To narrow down the possible solutions for the temperature gradient and compare the findings with \cite{Wyttenbach2015} and \cite{Pino2018}, we set $\NaX$ to $0$, corresponding to solar sodium abundance and a base pressure $P_0 = 10$~bar. Removing the degeneracy leads to a better convergence for the temperature gradient (see Figure \ref{fig:gradientposterior2}). Plotting our retrieved TP profiles (isothermal and temperature gradient) shows broad agreement with the work of \cite{Wyttenbach2015} and \cite{Huang2017} (see Figure \ref{fig:TPprofiles}). Work by \cite{Huang2017} has shown that an incremental fit of isothermal profiles as in \cite{Wyttenbach2015} underestimates the temperature in the higher layers of the atmosphere for atmospheres with rapidly increasing temperatures. Our work, which uses a temperature gradient, is in broad agreement with this conclusion, as our gradient is consistently on the higher temperature end of \cite{Wyttenbach2015} (see Figure \ref{fig:TPprofiles}) and encloses results from \cite{Huang2017} at the higher range of retrieved values. 

From the Bayesian evidence it is clear that there is no difference for this dataset of HD~189733b whether a isothermal profile or a temperature gradient is used, therefore the wind patterns are applied with an isothermal temperature profile as the base. 

\subsubsection{Isothermal profiles with added wind pattern}

\begin{figure*}[htb!]
\resizebox{\textwidth}{!}{\includegraphics[trim=0.0cm 12.0cm 0.0cm 0.0cm]{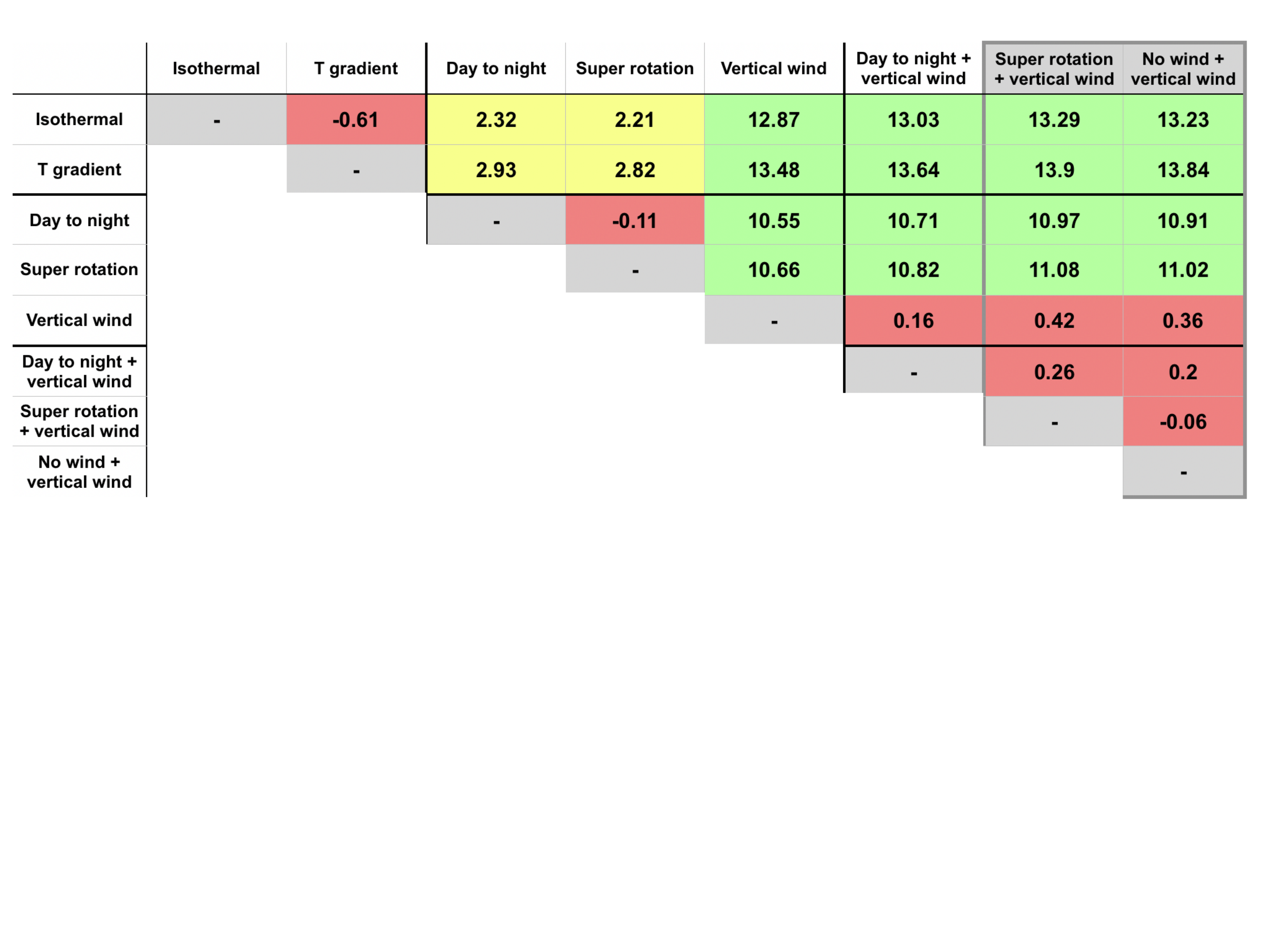}}
	\caption{Difference in Bayesian evidence from Table \ref{table:HD189733bcomparison}. The top horizontal models are compared to each other, where dark red means no evidence, yellow moderate evidence and light green strong evidence. All wind patterns that include vertical wind are strongly favoured over all models, the horizontal wind patterns show some improvement to no winds and we cannot distinguish between an isothermal profile or a temperature gradient. }
	\label{fig:retrievalresults}
\end{figure*}

When applying the additional basic wind profiles of a constant day-to-night side wind, a constant super-rotational wind or a constant vertical upward wind, Figure \ref{fig:retrievalresults} shows that all three are better fits than the profiles without any additional winds, with moderate evidence for the day-to-night and super-rotational wind and strong evidence for the vertical wind pattern. All three wind patterns converge to temperatures and NaX parameters within $1\sigma$ of each other and all three wind patterns converge clearly towards a value for the velocity, with the posterior as a Gaussian distribution. In the best-fit case of vertical upward winds, this means a mean isothermal temperature of $T(\mathrm{mean})=3426^{+316}_{-374}$~K, NaX fitting the continuum at $\mathrm{NaX}(\mathrm{mean})=-1.44^{+0.45}_{-0.33}$ and a vertical wind of $|\vec{v}_{\mathrm{ver}}(\mathrm{mean})|=40.55^{+3.82}_{-3.59}~\kms$. The vertical velocity is surprisingly large (see Section \ref{sec:implications}) and close to the escape velocity of HD~189733b of $|\vec{v}_{\mathrm{escape}}|=42.1~\kms$, which suggests that more complex wind patterns combined with a vertical upward velocity in the upper layers of the atmosphere could provide an even better fit to the data.
Based on this result, we implemented combinations of wind patterns. Given current literature favouring rotational wind patterns for high pressures (up to $10^-6$~bar) over vertical wind patterns, we split the atmosphere into two layers. A lower part and an upper part with different wind patterns each, where the split retrieved according to Equation \ref{eq:pswitch}.
When splitting the atmosphere into the lower and upper atmosphere, the Bayesian evidence suggests a significantly better fit than any of the other, more simple models, except the model of the vertical wind pattern throughout the atmosphere (see green cells in Figure \ref{fig:retrievalresults}). However, the posterior for the lower atmosphere day-to-night side model (see Figure \ref{fig:dtn_verposterior}) shows that the retrieval converges towards both the solution with only a vertical velocity in the atmosphere (h towards 0, the planet surface) and the split atmosphere solution (see best fit in blue in Figure \ref{fig:dtn_verposterior}), which subsequently generates a tail in the day-to-night side velocity distribution towards higher velocities. The best-fit is nonetheless found in the local likelihood minimum for the lower day-to-night side velocity pattern with $h=-6$. Assuming a super-rotational pattern in the lower atmosphere shows the same behaviour in the posterior distribution (see Figure \ref{fig:srot_verposterior}) with the best-fit still for a split into lower and higher atmosphere at $h=-5$ and a wide spread of possible wind velocities in the lower atmosphere $|\vec{v}_{\mathrm{srot}}|=8.89^{+7.03}_{-5.85}~\kms$. Since neither lower wind pattern conclusively converges towards a wind speed and atmospheric structure, it is likely that we cannot discriminate between the wind patterns in the lower atmosphere from the current quality of the data. A retrieval of no wind pattern in the lower atmosphere and a vertical outbound wind pattern in the upper atmosphere hardens this suspicion (see Figure \ref{fig:no_verposterior}). The retrieval converges to a split of the atmosphere at $h=-5.5$, with a tail for the lower atmosphere and similar values for the other retrieved parameters when compared to the same retrieval with the vertical wind pattern throughout the atmosphere (see Figure \ref{fig:vverticalposterior}). The vertical wind velocity remains high $|\vec{v}_{\mathrm{ver}}(\mathrm{mean})|=40.08^{+3.72}_{-3.40}~\kms$. Additionally, the Bayesian evidence for the three retrievals with vertical winds in the upper atmosphere and different wind patterns in the lower atmosphere are nearly the same, with a small preference for super rotation in the lower atmosphere. This means we cannot make any statements about the possible wind patterns' strength or shape in the lower atmosphere from the here presented dataset.

\begin{figure*}[htb!]
\resizebox{\textwidth}{!}{\includegraphics[trim=0.0cm 5.4cm 0.0cm 5.4cm]{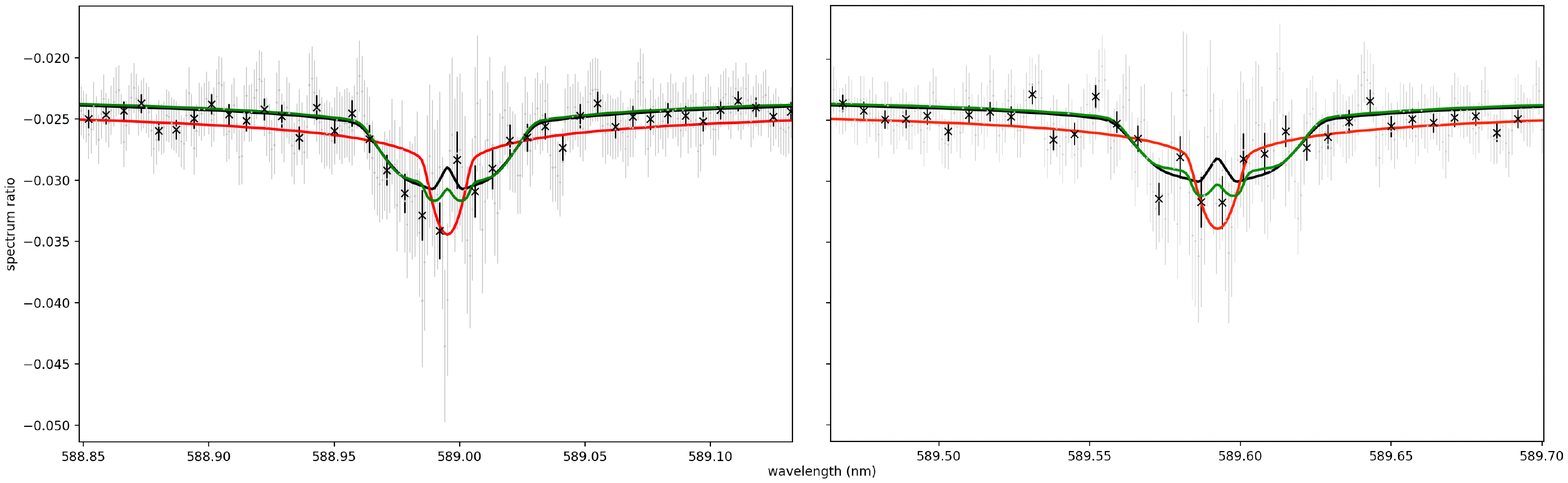}}
	\caption{Spectral ratio of the hot Jupiter HD189733b for the sodium D2 and D1 line. In grey the data obtained with the HARPS spectrograph at the ESO 3.6m telescope, in black the binned data for visibility. The red line shows the isothermal base model, compatible with findings from \cite{Wyttenbach2015}, and the black line the best-fit with super-rotation in the lower atmosphere and vertical winds in the upper atmosphere. The green line shows the equally good fit of the model with no wind for pressures higher than $h=-6$ and solely vertical winds farther up in the atmosphere. The "bump" for both the black and green fits stems from the parts of the atmosphere where the LOS is orthogonal to the wind direction and the profile is not Doppler shifted. The parameters used to generate the fits are indicated as the blue lines (best-fit) in Figures \ref{fig:isoposterior}, \ref{fig:srot_verposterior} and \ref{fig:no_verposterior} respectively.}
	\label{fig:bestfit}
\end{figure*}

A plot of the data with the isothermal base model and the two best-fit, and therefore preferred, models (no lower winds with higher vertical outbound wind and super-rotation in the lower atmosphere with vertical winds in the upper atmosphere) is shown in Figure \ref{fig:bestfit}.

\begin{table*}
\caption{Comparison of the different models. The base model to calculate $|\ln\mathcal{B}_{01}|$ is the isothermal model with no added wind patterns. The comparison stems from the Jeffrey's scale in Table \ref{table:Bayesianoverview}.}
\label{table:HD189733bcomparison}
\centering
\begin{tabular}{l c c l}
\hline
\hline
Model & $|\ln\mathcal{Z}|$   &  $|\ln\mathcal{B}_{01}|$ & Strength of evidence    \\
\hline
isothermal  &  $726.92\pm0.14$   & -  & - \\
T gradient  &  $726.31\pm0.13$   & $-0.31$  & No evidence \\
isothermal with day-to-night side wind  &  $729.24\pm0.17$   & $2.93$  & Moderate evidence \\
isothermal with super-rotational wind  &  $729.13\pm0.17$   & $2.82$  & Moderate evidence \\
isothermal with vertical wind  &  $739.79\pm0.17$   & $12.87$  & Strong evidence \\
\hline
isothermal with day-to-night side lower and vertical higher wind  &  $739.95\pm0.2$   & $13.0$  & Strong evidence \\
isothermal with no lower and vertical higher wind  &  $740.15\pm0.13$   & $13.20$  & Strong evidence \\
isothermal with super-rotational lower and vertical higher wind  &  $740.21\pm0.2$   & $13.26$  & Strong evidence \\

   \hline
\end{tabular}
\end{table*}

\subsection{Implications}
\label{sec:implications}

We explored an isothermal profile and a temperature gradient to fit the data, with different wind patterns added to the base model as described in Section \ref{sec:wind}. We found no evidence that a temperature gradient provides a better fit than the isothermal base model with the current data (see Table \ref{table:HD189733bcomparison} and Section \ref{sec:resultsHD189}) and use the isothermal profile as the base for all models with added winds. The temperature gradient for the best-fit model is $0.10^{+0.23}_{-0.22}$~K~km$^{-1}$, which is in agreement with the value found in \cite{Wyttenbach2015} and predicted by \cite{Heng2015}, but does not allow for a validation of these results due to the high errorbars. The retrieved value for the isothermal profile is in agreement with the top of the atmosphere results from \cite{Wyttenbach2015} as expected given that we retrieve on the line center, the same part used in \cite{Wyttenbach2015} for the upper atmosphere. Both the isothermal and the temperature gradient are shown in Figure \ref{fig:TPprofiles}, overplotted with their $1\sigma$ envelopes, together with the values from \cite{Wyttenbach2015} in black and the best-fit from \cite{Huang2017} in red. The temperature at zero altitude was taken from the continuum retrieval and are in agreement with results from Rayleigh scattering modelling in \cite{Lecavelier2008} for HD~189733b.

Adding a day-to-night side wind to the isothermal base profile generates a better fit (moderate evidence) just as a super-rotational wind profile throughout the atmosphere with a wind velocity of $3.8^{+1.51}_{-1.55}~\kms$, compatible with results from \cite{Salz2018}. The strongest evidence measured against the isothermal base model is found when employing a vertical wind profile from the surface of the atmosphere to the highest probed layer. However, the wind speed for the best-fit is $|\vec{v}_{\mathrm{ver}}|=40.50^{+3.82}_{-3.59}~\kms$, speeds unlikely to exist in the lower layers of the atmosphere and in contradiction with predictions from \cite{Salz2018, Flowers2019}. 

For a super-rotational wind in the lower atmosphere and a vertical wind in the upper atmosphere, the best-fit gives a lower wind speed of $3.0~\kms$, which changes to a best-fit vertical outwards bound wind of $38.90~\kms$ (within one sigma of the other retrieved vertical wind values), switching at an altitude of $10^{-6}$~bar (corresponding to h=-5 and NaX=-1). This is roughly the same relative altitude as the atmospheric layers studied by \cite{Salz2018,Flowers2019} and therefore compatible with their findings of a super-rotational wind of $0 - 4.5~\kms$ for altitudes lower than $10^{-5}$~bar (see Figure 2 in \cite{Flowers2019}). However, this comparison is only on an order of magnitude level, given the uncertainty of the here presented retrieval values for the lower atmosphere.
 
The retrieved vertical wind speed in the upper atmosphere is associated with a scenario relying on multiple assumptions and simplifications, and suggests that additional mechanisms determine the shape the sodium line in HD~189733b. Due to the geometrical simplifications used in our forward model, we did not implement the rotation of the tidally locked planet ($v_{rot}=2.7~\kms$). Therefore, we present the wind speed only as an upper boundary of the wind speeds needed to explain the broadening. 

A full physical modelling as to the origin of this wind is beyond the scope of this paper, yet we would like to provide a possible scenario on the order of magnitude level leading to comparable wind speeds in planets similar to HD~189733b. 
We propose a super-rotational jet in the lower part of the atmosphere with wind speeds of the order of magnitude of a few $\kms$ based on \cite{Salz2018}. While some work has been done on vertical shear in the super-rotational jet of HD~189733b \citep{Brogi2016}, it is not sufficient to introduce an acceleration ($\vec{a}$) capable of propelling atoms to the velocities proposed in this paper. However, if ions at the speeds found in the super-rotational jet are subjected to a magnetic field in the equatorial region, the Lorenz force creates an upwards movement of the ions. Literature on the transition zones between different atmospheric layers in Jupiter and Jupiter-like worlds is sparse, but it can be assumed that these zones are slim compared to the overall structures in the atmosphere \cite{Robinson2014}. 
The ionisation fraction $Na^+/Na$ is computed with the Saha equation, where we use the number density of free electrons computed following a procedure from \cite{Molliere2017}. 
We found that only a small fraction remains as neutral atoms. Assuming the Lorenz force is the potential source of these large velocities, we calculated the necessary magnetic field to propel a sodium ion to the observed speeds at the base of the upper atmosphere via:

\begin{equation}
m \cdot \vec{a}_{\mathrm{ver}} = q \cdot \vec{v_{\mathrm{rot}}} \times \vec{B}
\end{equation}

where $\vec{a}_{\mathrm{ver}}$ is the necessary acceleration to propel particles to the retrieved vertical wind speed over the slim transitional region between atmospheric layers (order of magnitude $\sim1$~km), and $\vec{v_{\mathrm{rot}}}$ is the velocity of particles inside the super-rotational jet.
This leads to an order of magnitude for magnetic field strength ($\sim50~G$) orders of magnitude higher than the predicted fiels strength for planets with rotation periods of $\sim2-4$ days \cite{Cauley2019}. However, the needed magnetic field strength is comparable at the order of magnitude level with the magnetic field strength measured for HD~189733b \citep{Cauley2019}. The ions recombine rapidly to neutral species, which are then at the accelerated speed of the former ions. Furthermore, the remaining ions could also drag the neutral species upwards. In this context, it is worth mentioning that the very top of the sodium line probes high altitudes, where collisions between the sodium atoms become less and less frequent, which favours a constant speed of the particles compared to a deceleration.

Additional effects that could have impacts on the sodium line broadening and change the results from our analysis are ionisation due to the close proximity to the host star and non-LTE effects. In the case of ionisation, extra ionisation sources potentially deplete the upper layers of the atmosphere of neutral sodium. The dearth of sodium creates a more shallow line center, while the line wings keep their shape. This can be mistaken for a full sodium line with broad wings instead, creating a degeneracy between broadened line wings or strong ionisation as possible physical explanations of the line shape \citep{Lothringer2019}.
Non-LTE effects might also influence the line shape, however these effects cannot be distinguished with the current data quality and might have a smaller contribution \citep{Fisher2019}.

\cite{Lecavelier2010, Lecavelier2012, Bourrier2013}  studied HD~189733b in Lyman~$\alpha$ and indicated an expanding exosphere. \cite{Vidal-Madjar2003,Lammer2003} has also shown, that the planet is evaporating and that it is thought to arise from an expanding thermosphere, reaching high velocities in the upper parts. Considering the heights we probe (see Figure \ref{fig:TPprofiles}), it is possible we reach these layers and see sodium atoms propelled to these velocities close to the escape velocity.

\section{Conclusions}
 
We introduced the MERC code - a new tool to combine 1D atmospheric forward models with pseudo 3D wind patterns together with a rigorous nested sampling algorithm. The pseudo 3D wind is created by calculating the Doppler broadening for each cell in one or more atmospheric slices and rotating them to create the full atmosphere. From this model atmosphere, a model transmission spectrum is calculated and subsequently compared to the real data via the multi-nested sampling module. To achieve a correct fitting for wind strengths and patterns, a separation of the continuum of the data from the line cores was introduced. We showed the self consistency of MERC by running it on simulated data, highlighting its ability to not only correctly retrieve the parameter values but also to distinguish between models successfully. 
We then applied our technique to the well studied exoplanet HD~189733b on the sodium doublet first published in \cite{Wyttenbach2015} and corrected for the RM effect. We tested an isothermal temperature profile and a temperature gradient, highlighting that both solutions provide satisfactory fits. We then explored three additional wind profiles: a day-to-night side wind, a super-rotational wind and a vertical upward wind. Although the vertical wind throughout the atmosphere was preferred, we expanded this result by splitting the atmosphere in a lower and upper atmospheric layer. Testing the established wind patterns in the lower atmosphere with an additional vertical wind in the upper atmosphere showed that the current data quality does not allow us to distinguish wind patterns in the lower atmosphere. However, the solution with no assumptions as to the wind profile in the lower atmosphere and a vertical wind in the upper atmosphere was clearly retrieved as the best solution. Our best solution has a mean temperature of $T(\mathrm{mean}) = 3412^{+347}_{-432}$~K and a vertical upward wind of $|\vec{v}_{\mathrm{ver}}(\mathrm{mean})|=40.08^{+3.72}_{-3.40}~\kms$ in the higher atmosphere up to the thermosphere.

The broadened sodium line in HD~189733b can thus be explained by strong winds in the upper layers of the atmospheres up to the thermosphere, possibly propelled by strong magnetic fields. 
 
With the next generation of spectrographs (e.g. ESPRESSO), we hope to  distinguish between different models in the lower atmosphere and classify models by their Bayesian evidence at various altitudes above the planets surface, as well as implement other possible broadening sources in MERC.


\begin{acknowledgements}
This project has received funding from the European Research Council (ERC) under the European Union's Horizon 2020 research and innovation programme (project {\sc Four Aces}; grant agreement No. 724427).
This work has been carried out within the frame of the National Centre for Competence in Research `PlanetS' supported by the Swiss National Science Foundation (SNSF). L.P. acknowledges that the research leading to these results has received funding from the European Research Council (ERC) under the European Union's Horizon 2020 research and innovation program (grant agreement no. 679633; Exo-Atmos). We thank N. Casasayas-Barris for their assistance with the RM benchmarking and L. Dos~Santos for their helpful comments. We would also like to show our gratitude to Kevin Heng for discussions and insights during the course of this research, that greatly improved the understanding of theoretical issues in atmospheric modelling.We thank P. Molli\`{e}re for the use of his code to calculate the ionisation fraction and the anonymous referee for their helpful comments and time.
\end{acknowledgements}
%
\bibliographystyle{aa} 
\bibliography{wind}
%

\begin{appendix}
\onecolumn
\section{Posterior distributions simulated data}
\label{app:sim}

\begin{figure*}[hbt]
\resizebox{\textwidth}{!}{\includegraphics[trim=4.0cm 0.0cm 4.0cm 0.0cm]{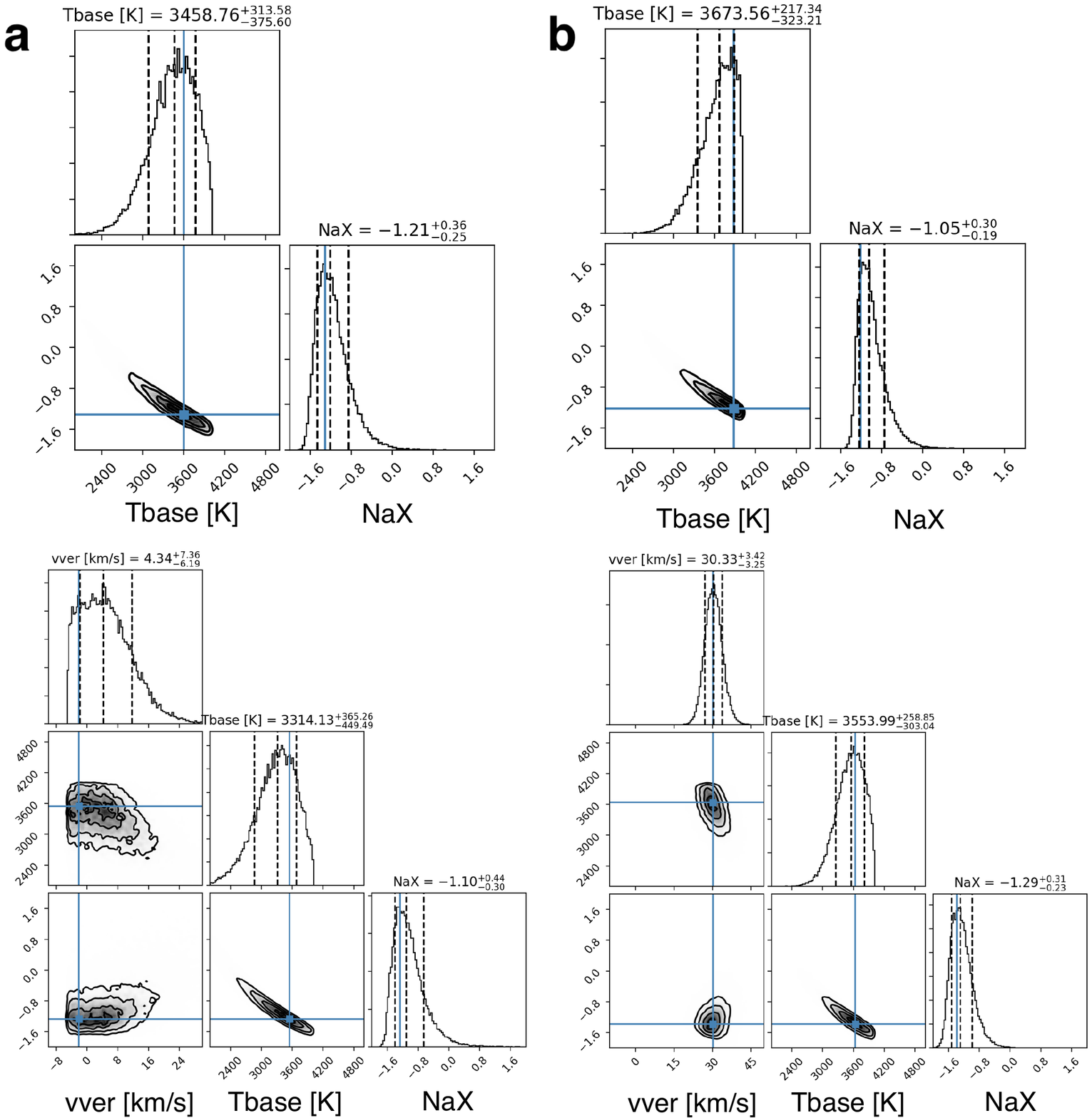}}
	\caption{Posterior distributions on the simulated data. Column \textbf{a} was retrieved on the isothermal simulated data with no additional winds. The top panel is the retrieval with an isothermal forward model, the bottom panel with an added vertical wind. Column \textbf{b} was retrieved on the simulated data from an isothermal temperature profile with added vertical upward winds. The top panel was retrieved with an isothermal forward model and the bottom panel with the added wind. All results were within expectations.}
	\label{fig:simposterior}
\end{figure*}

\begin{figure*}
\resizebox{\textwidth}{!}{\includegraphics[trim=4.0cm 4.9cm 4.0cm 4.9cm]{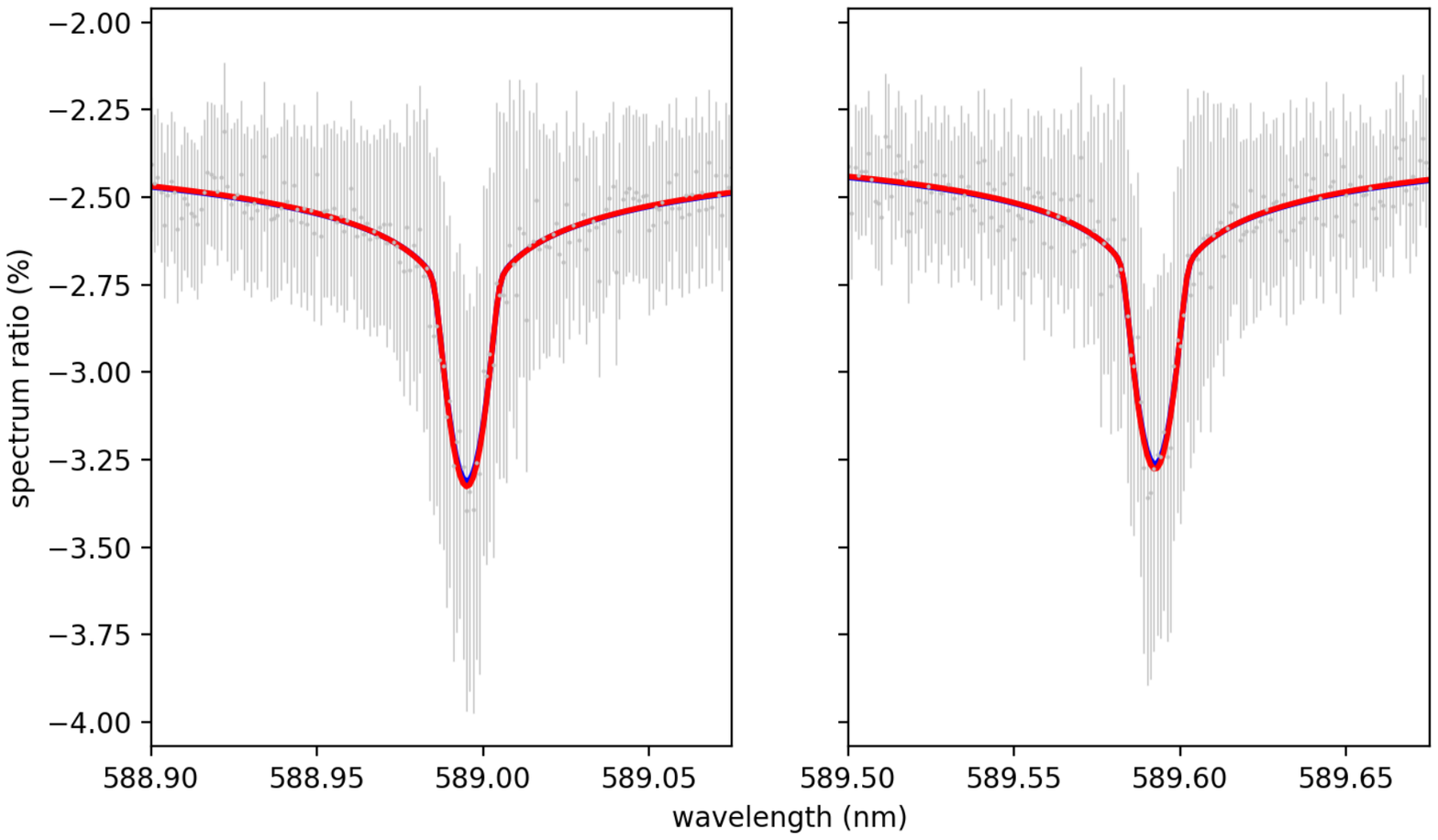}}
	\caption{Simulated data from an isothermal atmosphere in grey overplotted with the best-fit model. In red the isothermal model with no winds is shown and in blue the isothermal model with added vertical wind. Both models provide an identical fit.}
	\label{fig:simisoplot}
\end{figure*}

\begin{figure*}
\resizebox{\textwidth}{!}{\includegraphics[trim=2.0cm 9.0cm 2.0cm 9.1cm]{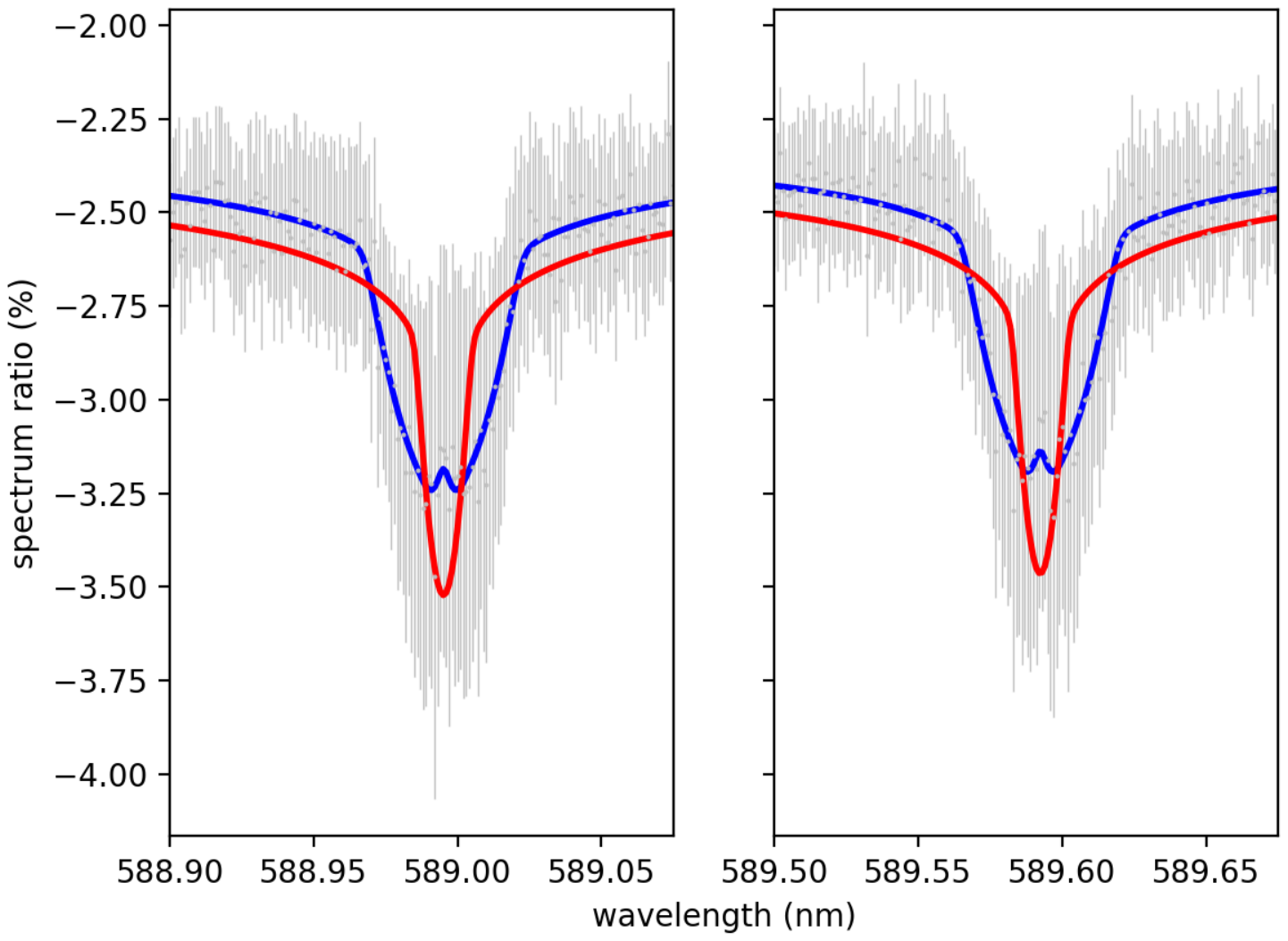}}
	\caption{Simulated data from an isothermal atmosphere with vertical upward winds added in grey overplotted with the best-fit model. In red the isothermal model with no winds is shown and in blue the isothermal model with added vertical wind.}
	\label{fig:simverplot}
\end{figure*}

\pagebreak
\section{Posterior distributions HD~189733b}
\label{app:HD189}

\begin{figure*}[hbt]
\resizebox{\textwidth}{!}{\includegraphics[trim=-2.0cm 5.5cm -9.0cm 5.5cm]{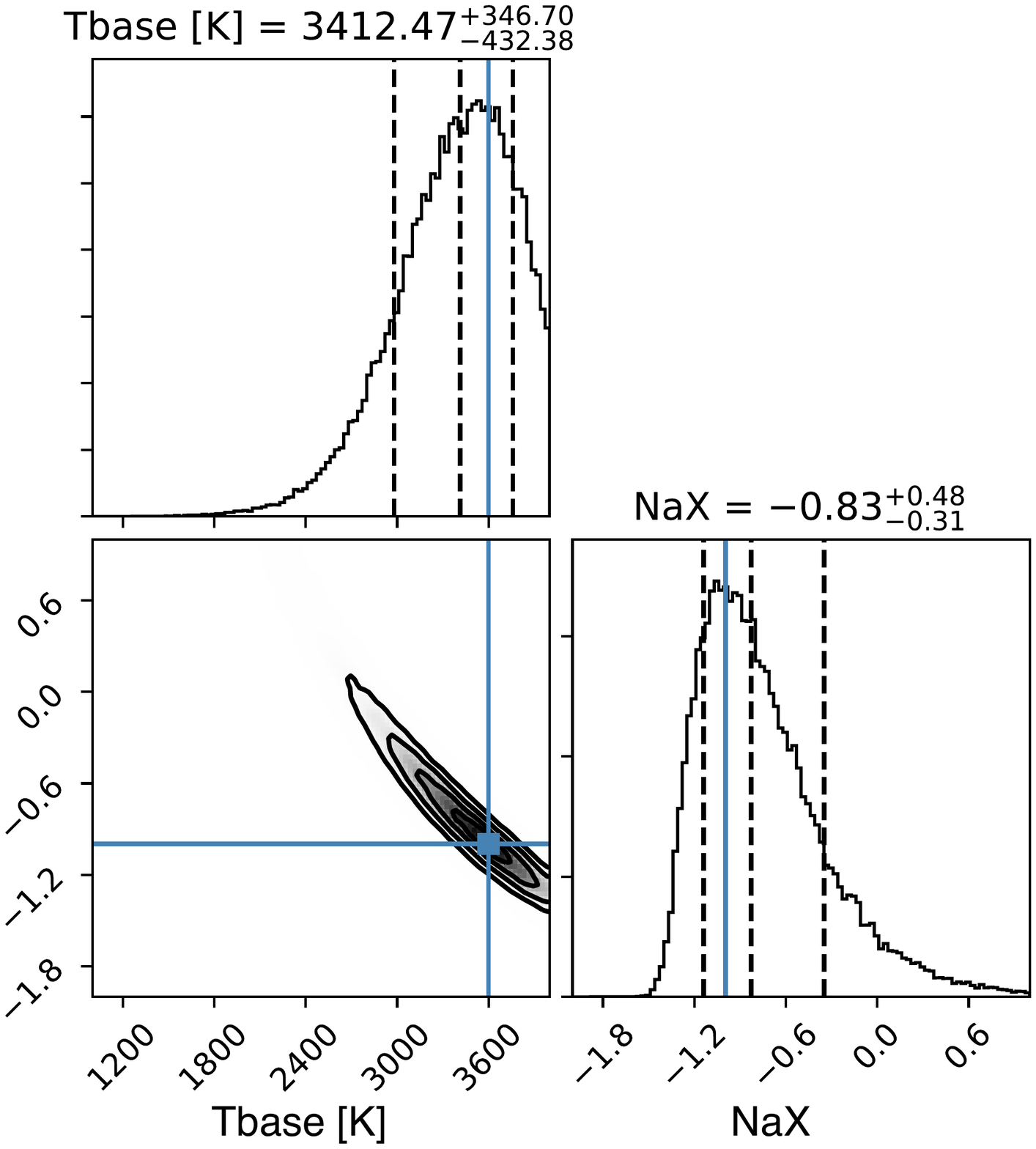}}
	\caption{Posterior distribution of isothermal line retrieval. This was used as the base model to compare all other scenarios to.}
	\label{fig:isoposterior}
\end{figure*}

\begin{figure*}[hbt]
\resizebox{\textwidth}{!}{\includegraphics[trim=-1.0cm 9.0cm -1.0cm 9.0cm]{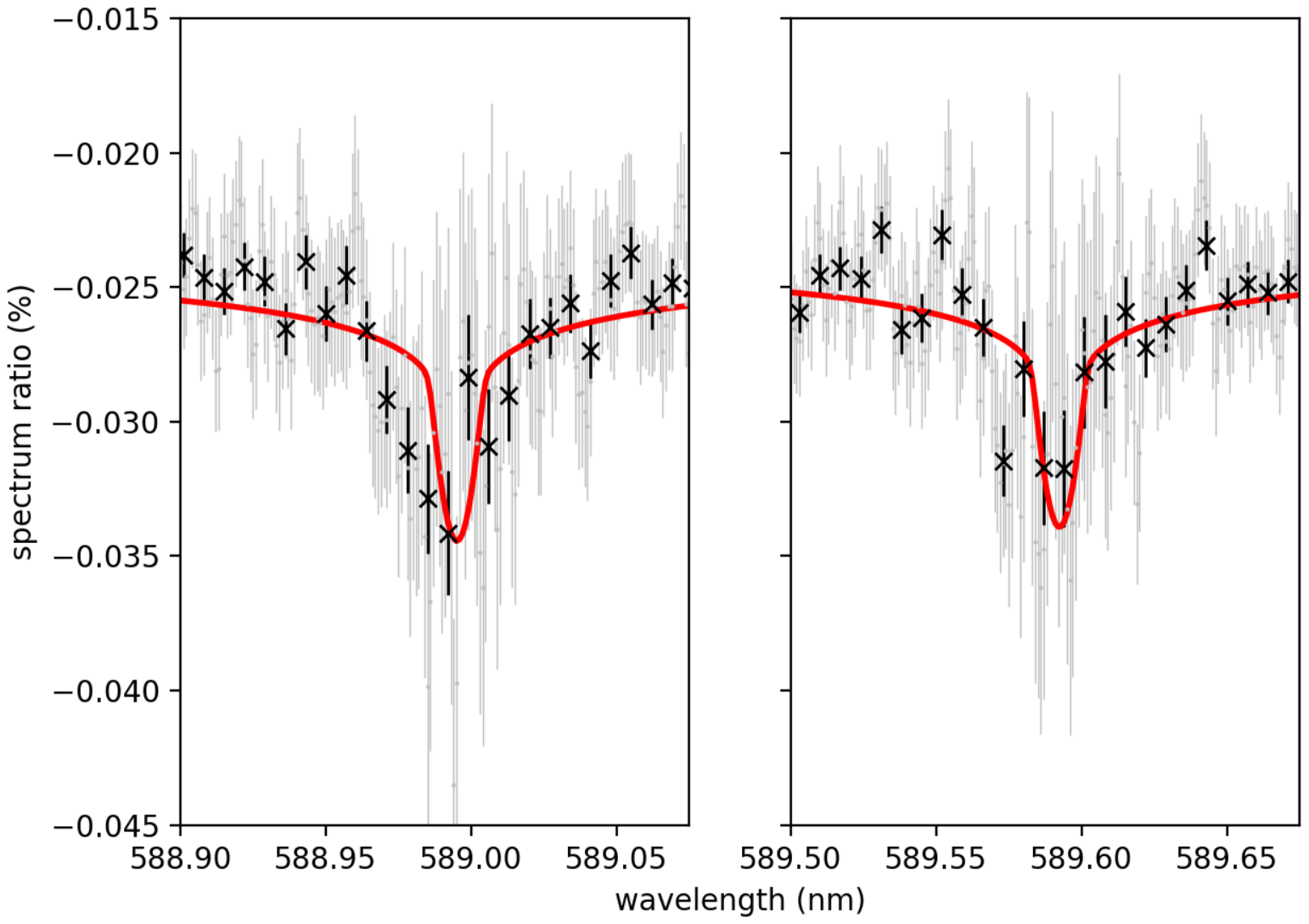}}
	\caption{Transmission spectrum of HD~189733b for both lines of the sodium doublet in grey, binned in black, with the best-fit retrieved by MERC in red. The best-fit was generated for the isothermal base model with no added winds and its best-fit parameters from the posterior distribution in Figure \ref{fig:isoposterior}.}
	\label{fig:isobest}
\end{figure*}

\begin{figure*}[htb!]
\resizebox{\textwidth}{!}{\includegraphics[trim=0.0cm 5.0cm -3.0cm 5.0cm]{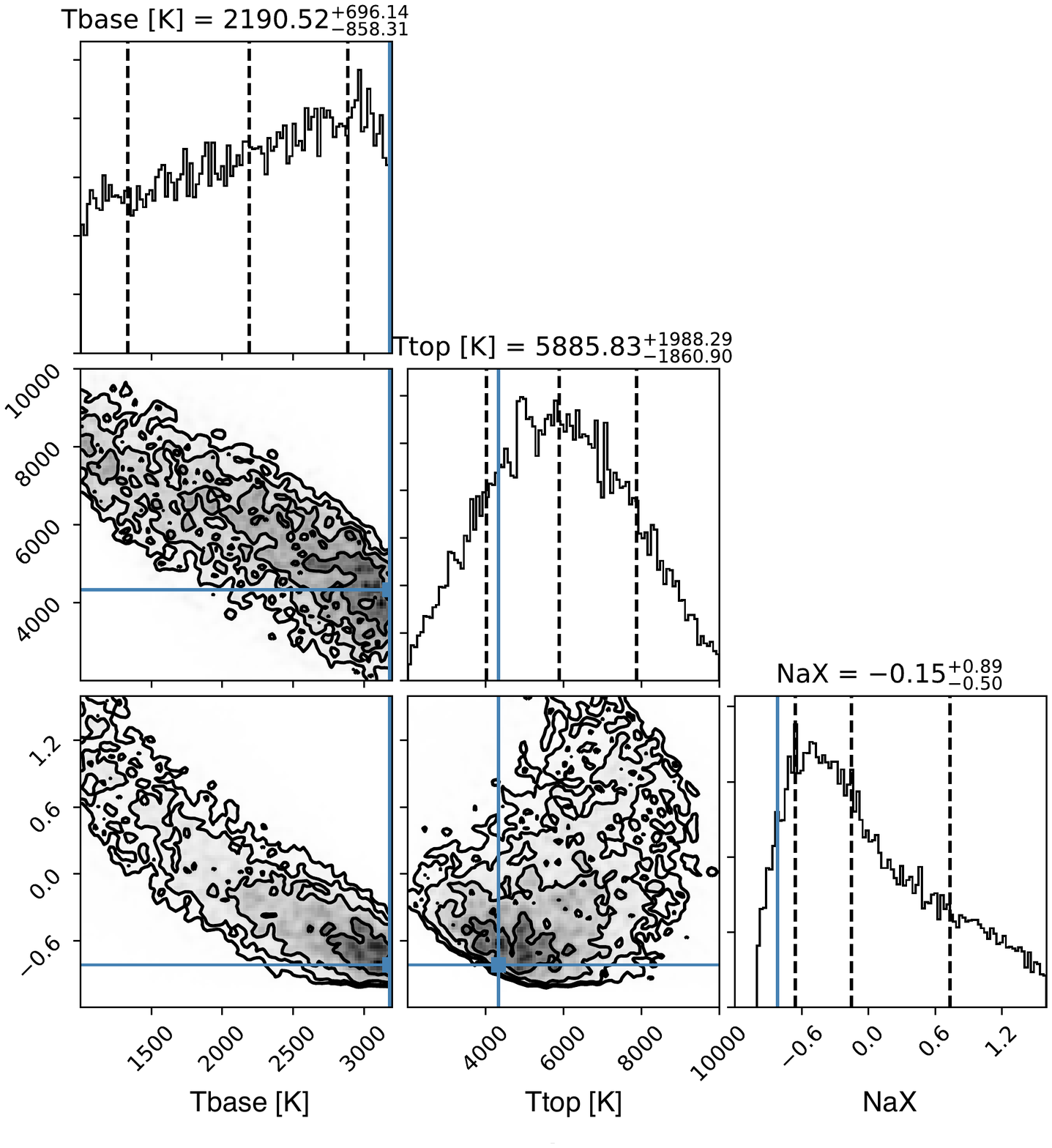}}
	\caption{Posterior distribution of temperature gradient line retrieval. T base is at the surface of the planet and T top at the top of the probed atmosphere. Due to the degeneracy between the three parameters, the convergence for this model is poor.}
	\label{fig:gradientposterior}
\end{figure*}

\begin{figure*}[htb!]
\resizebox{\textwidth}{!}{\includegraphics[trim=2.0cm 8.0cm -2.0cm 8.0cm]{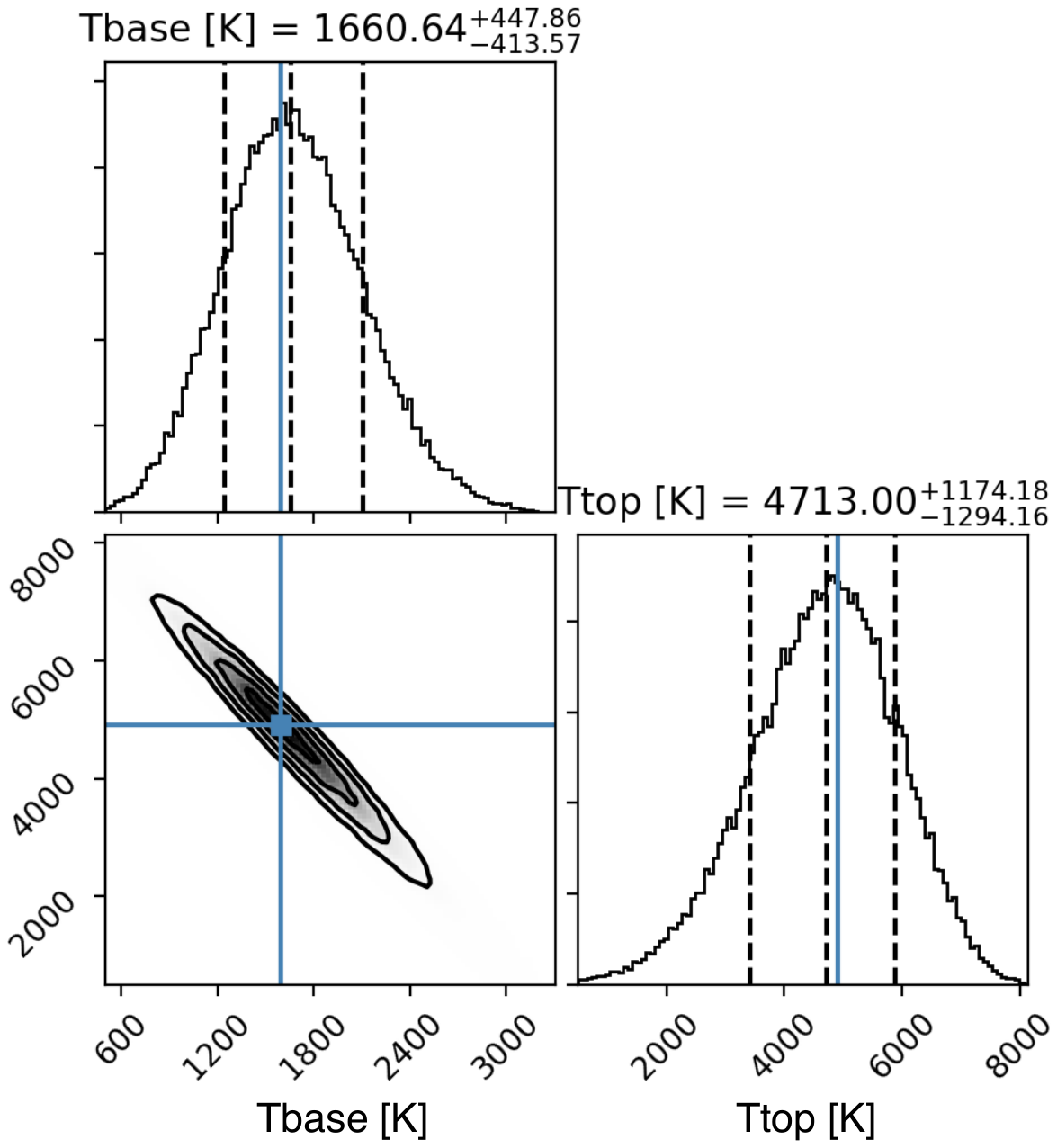}}
	\caption{Posterior distribution of temperature gradient line retrieval. T base is at the surface of the planet and T top at the top of the probed atmosphere. The continuum parameter $\NaX$ is set to $0$ to retrieve a converged temperature gradient and compare it to existing literature.}
	\label{fig:gradientposterior2}
\end{figure*}

\begin{figure*}[hbt]
\resizebox{\textwidth}{!}{\includegraphics[trim=-1.0cm 9.0cm -1.0cm 9.0cm]{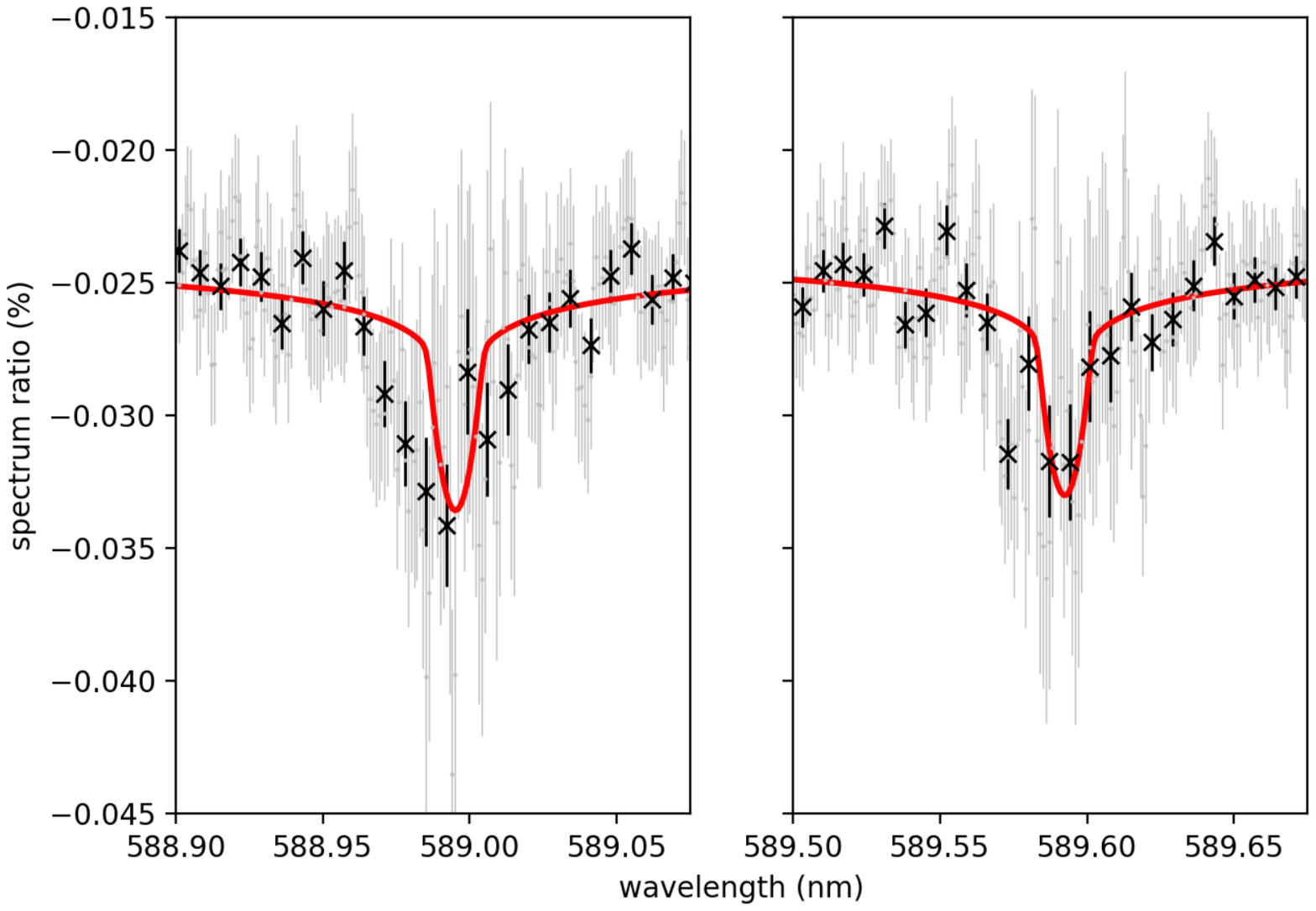}}
	\caption{Transmission spectrum of HD~189733b for both lines of the sodium doublet in grey, binned in black, with the best-fit retrieved by MERC in red. The best-fit was generated for the temperature gradient model with no added winds and its best-fit parameters from the posterior distribution in Figure \ref{fig:gradientposterior2}.}
	\label{fig:gradbest}
\end{figure*}

\begin{figure*}[htb!]
\resizebox{\textwidth}{!}{\includegraphics[trim=0.0cm 5.0cm -8.0cm 5.0cm]{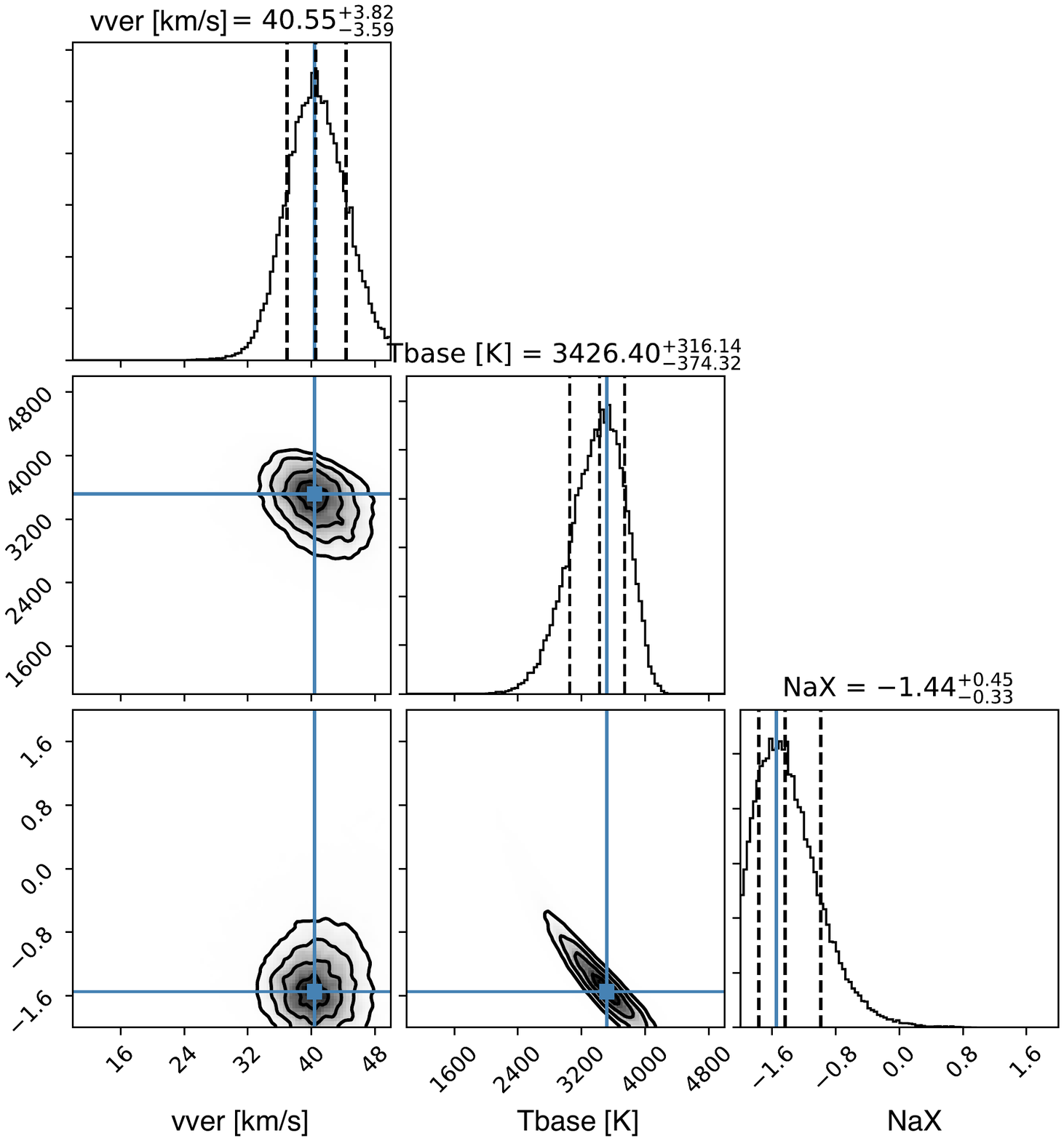}}
	\caption{Posterior distribution of isothermal line retrieval with an added vertical wind constant throughout the atmosphere.}
	\label{fig:vverticalposterior}
\end{figure*}

\begin{figure*}[hbt]
\resizebox{\textwidth}{!}{\includegraphics[trim=-1.0cm 9.0cm -1.0cm 9.0cm]{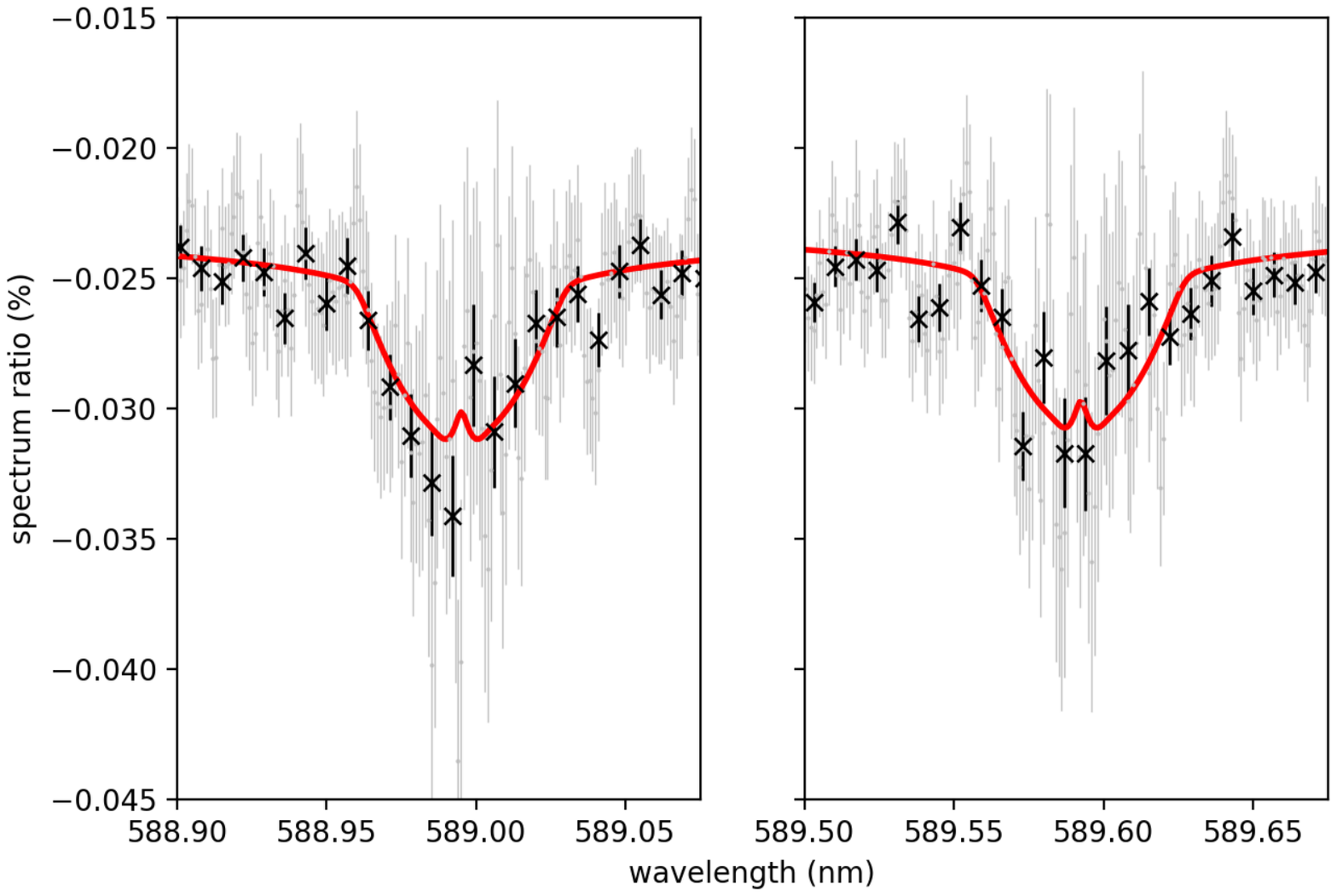}}
	\caption{Transmission spectrum of HD~189733b for both lines of the sodium doublet in grey, binned in black, with the best-fit retrieved by MERC in red. The best-fit was generated for the temperature gradient model with added vertical winds and its best-fit parameters from the posterior distribution in Figure \ref{fig:vverticalposterior}.}
	\label{fig:vverticalbest}
\end{figure*}

\begin{figure*}[htb!]
\resizebox{\textwidth}{!}{\includegraphics[trim=0.0cm 5.0cm -8.0cm 5.0cm]{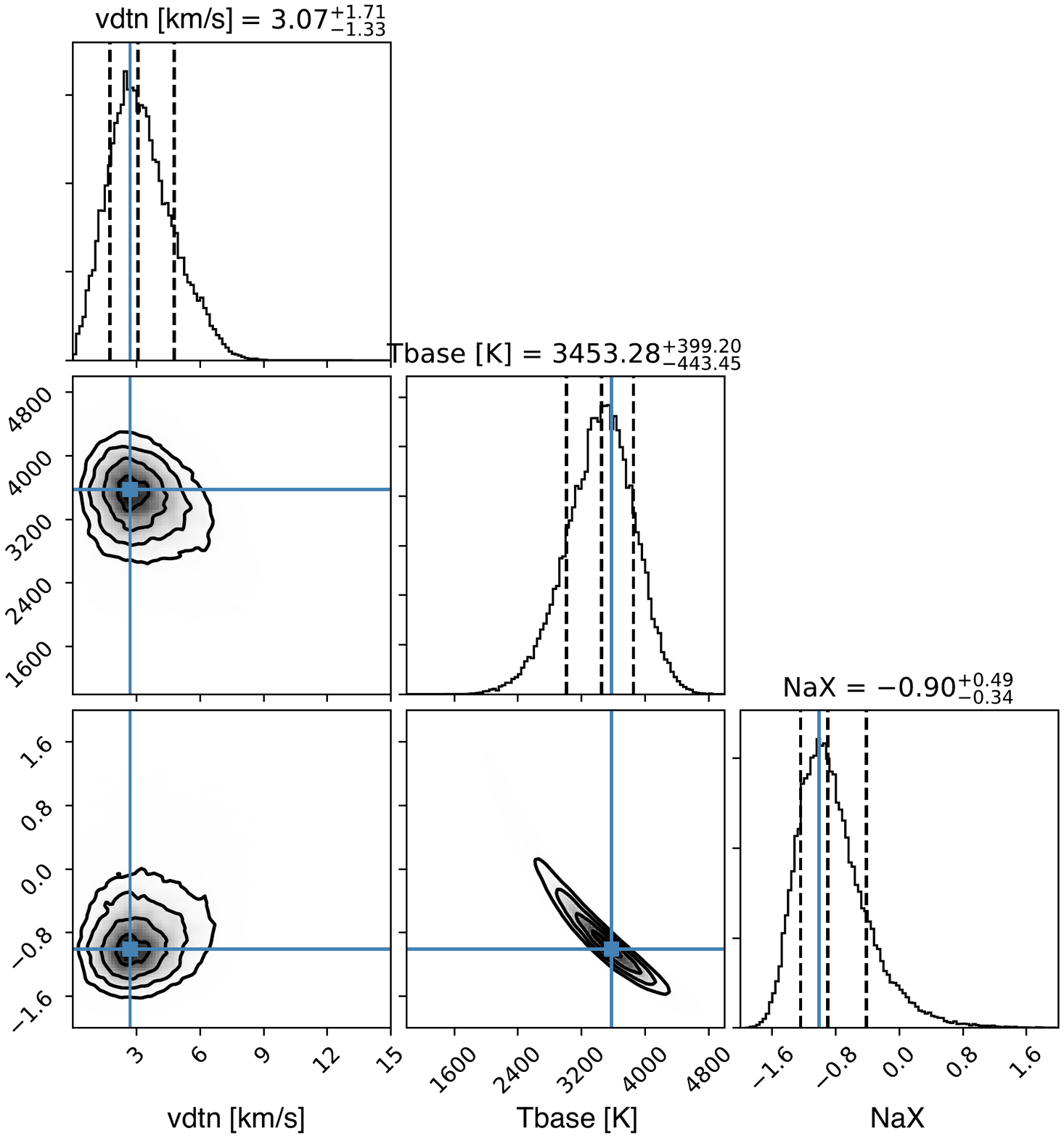}}
	\caption{Posterior distribution of isothermal line retrieval with an added day-to-night side wind constant throughout the atmosphere. }
	\label{fig:vdtnposterior}
\end{figure*}

\begin{figure*}[hbt]
\resizebox{\textwidth}{!}{\includegraphics[trim=-1.0cm 9.0cm -1.0cm 9.0cm]{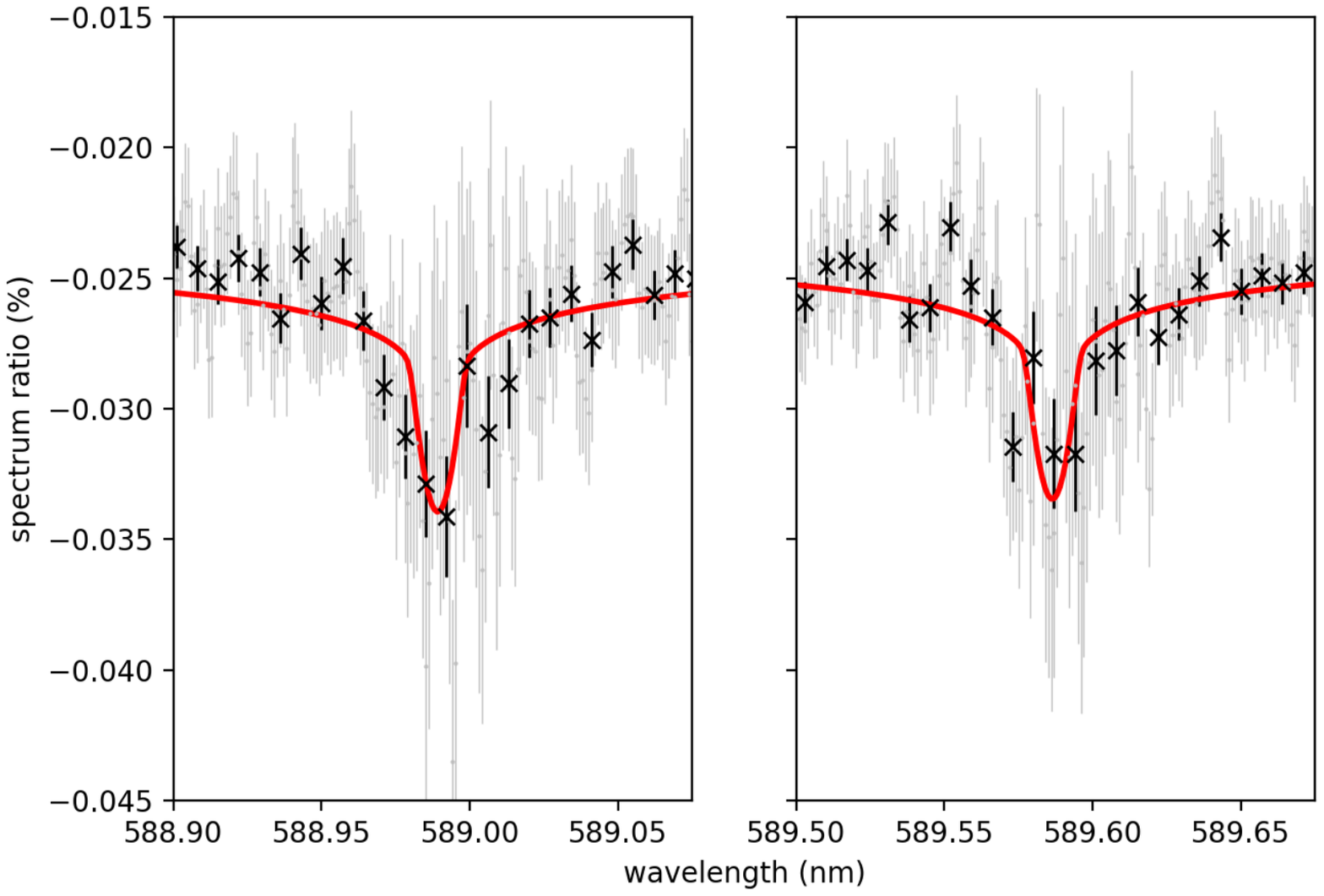}}
	\caption{Transmission spectrum of HD~189733b for both lines of the sodium doublet in grey, binned in black, with the best-fit retrieved by MERC in red. The best-fit was generated for the temperature gradient model with added day to night side winds and its best-fit parameters from the posterior distribution in Figure \ref{fig:vdtnposterior}.}
	\label{fig:vdtnbest}
\end{figure*}

\begin{figure*}[htb!]
\resizebox{\textwidth}{!}{\includegraphics[trim=0.0cm 5.0cm -8.0cm 5.0cm]{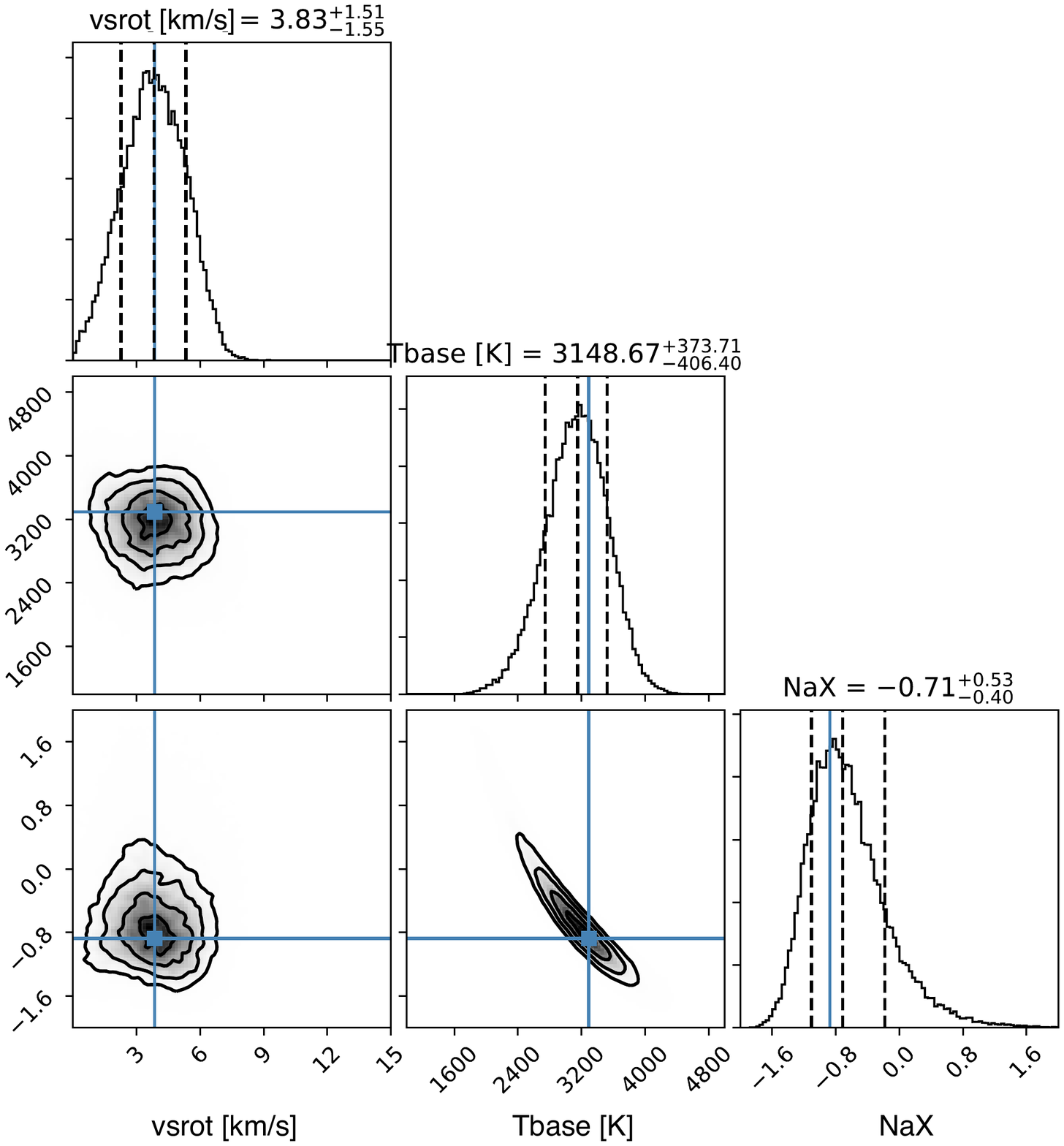}}
	\caption{Posterior distribution of isothermal line retrieval with an added super-rotational wind constant throughout the atmosphere. }
	\label{fig:superrotposterior}
\end{figure*}

\begin{figure*}[hbt]
\resizebox{\textwidth}{!}{\includegraphics[trim=-1.0cm 9.0cm -1.0cm 9.0cm]{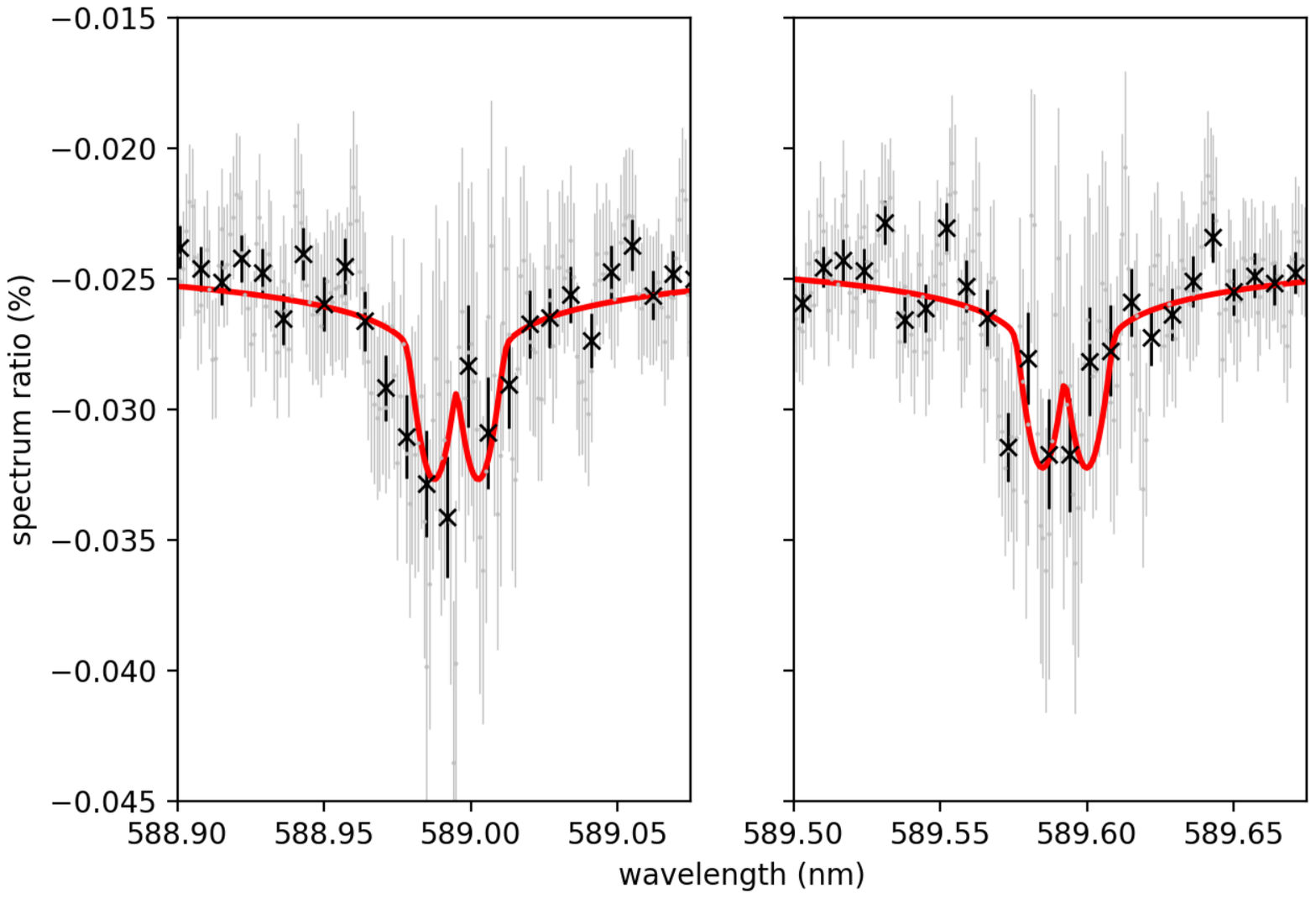}}
	\caption{Transmission spectrum of HD~189733b for both lines of the sodium doublet in grey, binned in black, with the best-fit retrieved by MERC in red. The best-fit was generated for the temperature gradient model with added super-rotational winds and its best-fit parameters from the posterior distribution in Figure \ref{fig:superrotposterior}.}
	\label{fig:vsrotbest}
\end{figure*}

\begin{figure*}[htb!]
\resizebox{\textwidth}{!}{\includegraphics[trim=1.0cm 2.0cm 2.0cm 2.0cm]{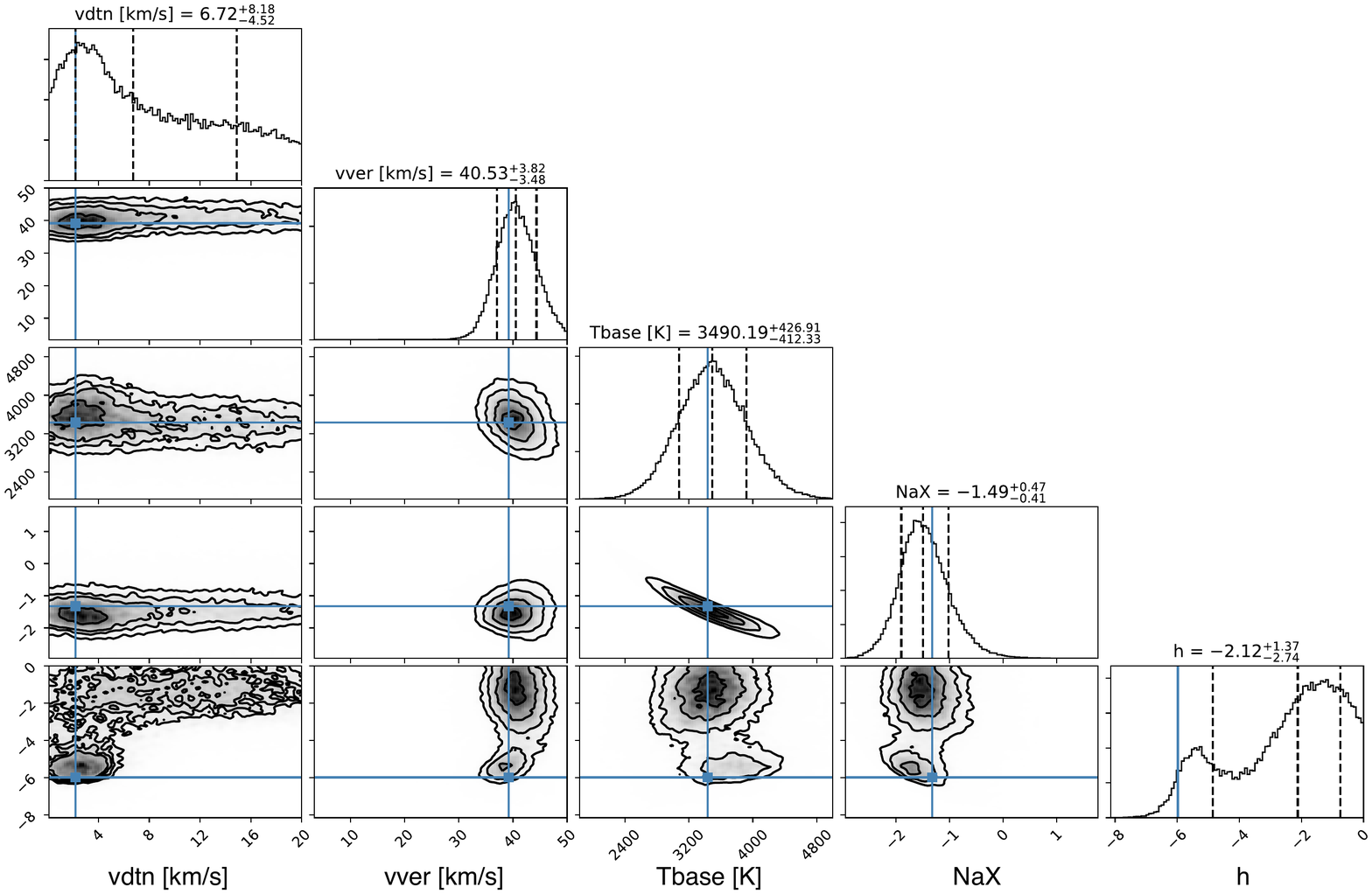}}
	\caption{Posterior distribution of isothermal line retrieval with an added day-to-night side wind in the lower atmosphere and a vertical wind in the upper atmosphere. The parameter on the right $h$ is then used to calculate the pressure where the velocity models are switched via Equation \ref{eq:pswitch}. Most notably, the retrieval is able to produce a solution with a day-to-night side wind in the upper layers of the atmosphere, but due to the data quality the solution with solely a vertical wind ($h$ closer to 0.) is degenerate (we would like to point out the two local maxima for $h$). }
	\label{fig:dtn_verposterior}
\end{figure*}

\begin{figure*}[hbt]
\resizebox{\textwidth}{!}{\includegraphics[trim=-1.0cm 9.0cm -1.0cm 9.0cm]{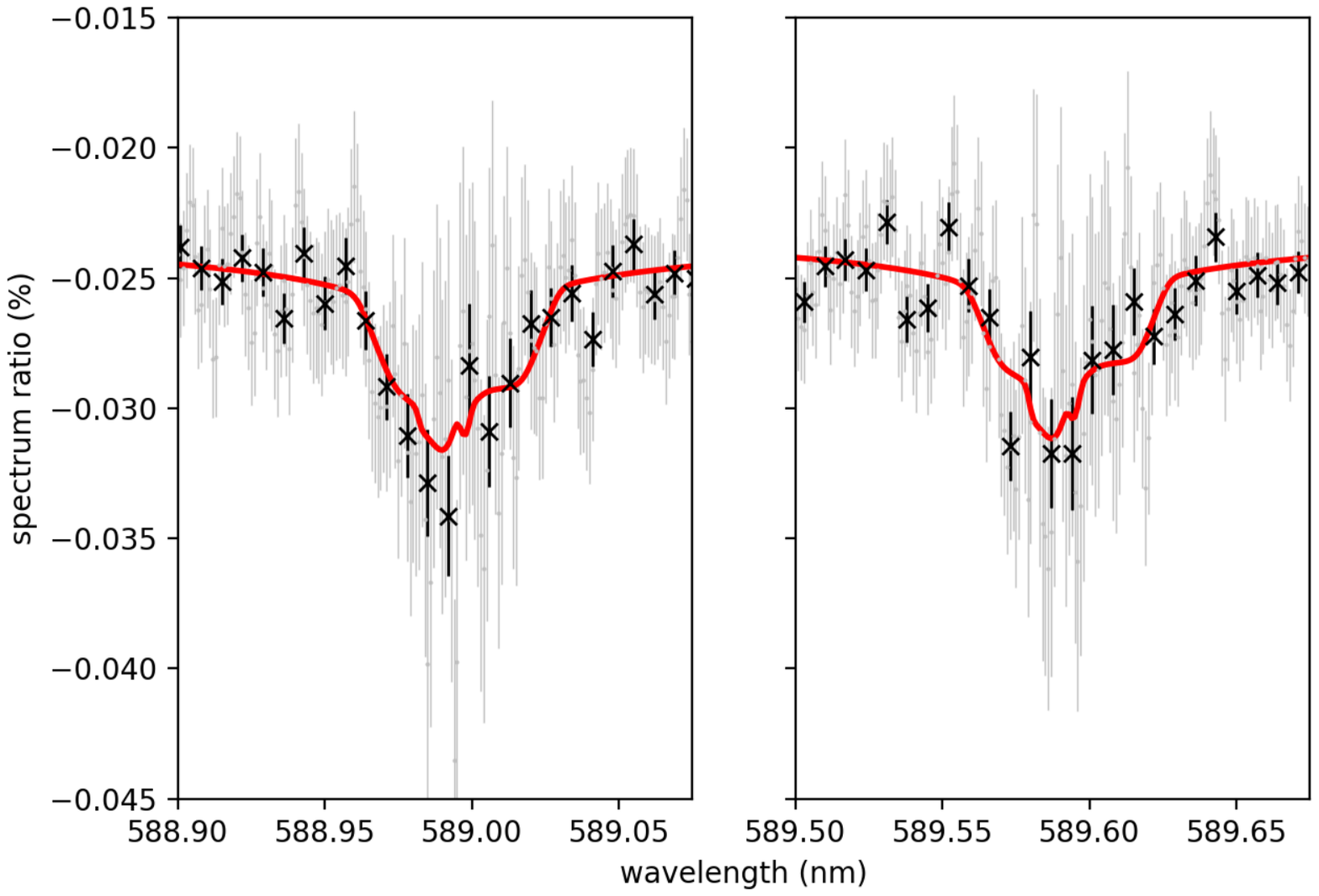}}
	\caption{Transmission spectrum of HD~189733b for both lines of the sodium doublet in grey, binned in black, with the best-fit retrieved by MERC in red. The best-fit was generated for the temperature gradient model with added day to night side winds in the lower atmosphere and vertical winds in the upper atmosphere with its best-fit parameters from the posterior distribution in Figure \ref{fig:dtn_verposterior}.}
	\label{fig:vdtnvverbest}
\end{figure*}

\begin{figure*}[htb!]
\resizebox{\textwidth}{!}{\includegraphics[trim=0.0cm 2.0cm 1.0cm 2.0cm]{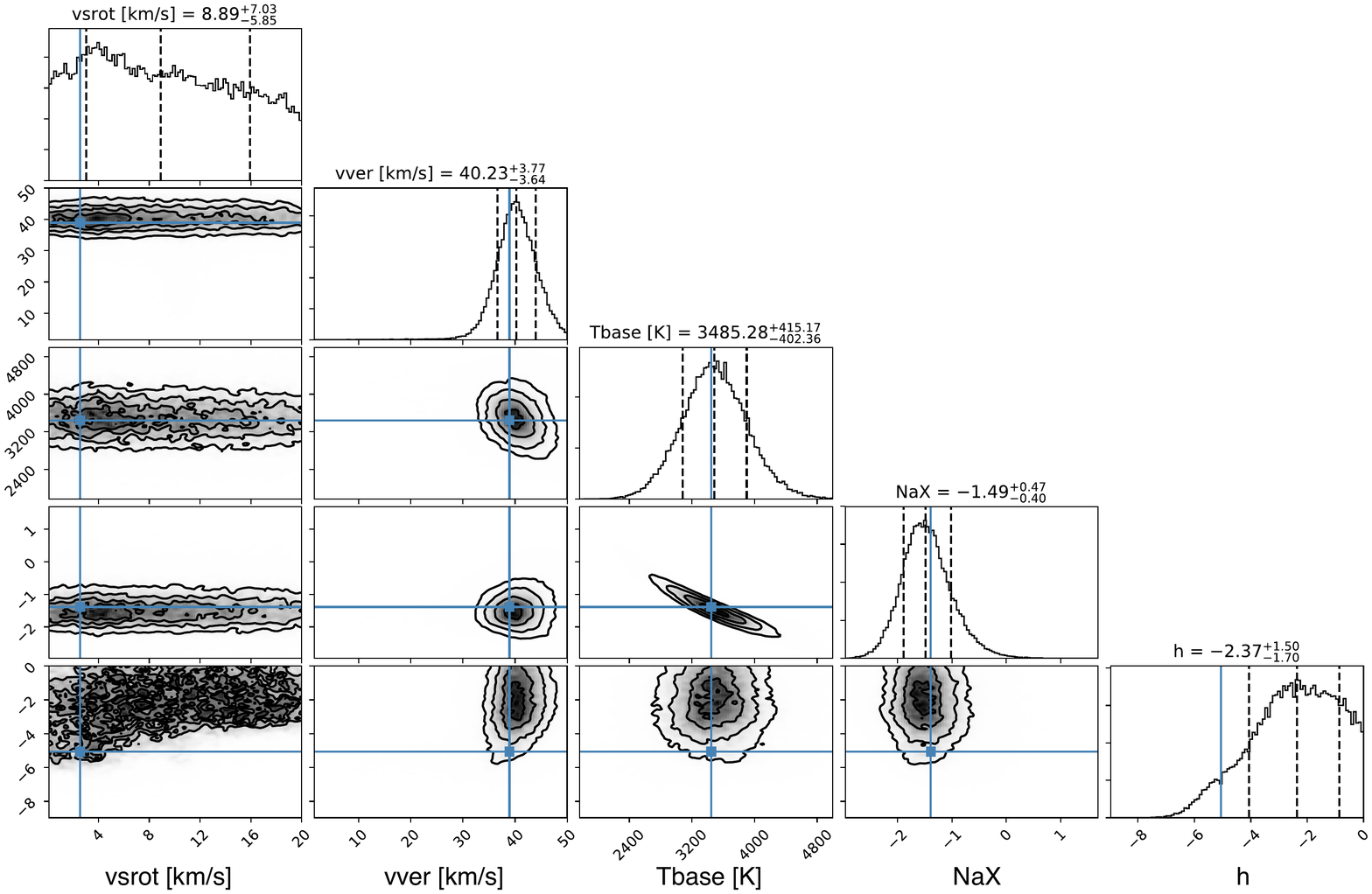}}
	\caption{Posterior distribution of isothermal line retrieval with an added super-rotational wind in the lower atmosphere and a vertical wind in the upper atmosphere. The parameter on the right $h$ is then used to calculate the pressure where the velocity models are switched via Equation \ref{eq:pswitch}. Just like the day-to-night side wind, the preferred solution of a super-rotational wind in the upper atmosphere ($h=-5$) is degenerate with a solely vertical wind ($h$ converging towards 0). }
	\label{fig:srot_verposterior}
\end{figure*}

\begin{figure*}[hbt]
\resizebox{\textwidth}{!}{\includegraphics[trim=-1.0cm 9.0cm -1.0cm 9.0cm]{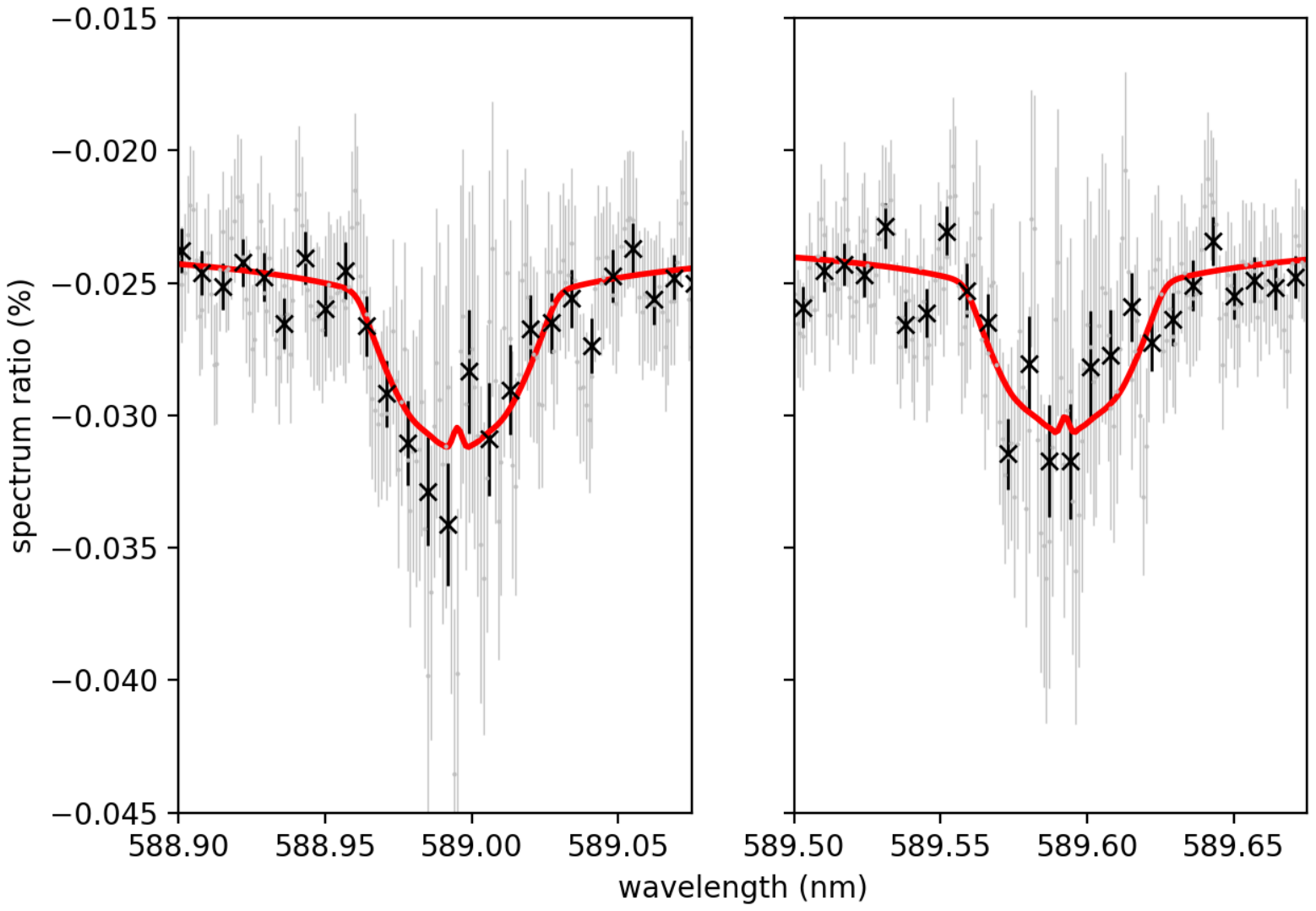}}
	\caption{Transmission spectrum of HD~189733b for both lines of the sodium doublet in grey, binned in black, with the best-fit retrieved by MERC in red. The best-fit was generated for the temperature gradient model with added super-rotational winds in the lower atmosphere and vertical winds in the upper atmosphere with its best-fit parameters from the posterior distribution in Figure \ref{fig:srot_verposterior}.}
	\label{fig:vsrotvverbest}
\end{figure*}

\begin{figure*}[htb!]
\resizebox{\textwidth}{!}{\includegraphics[trim=0.0cm 2.0cm 1.0cm 2.0cm]{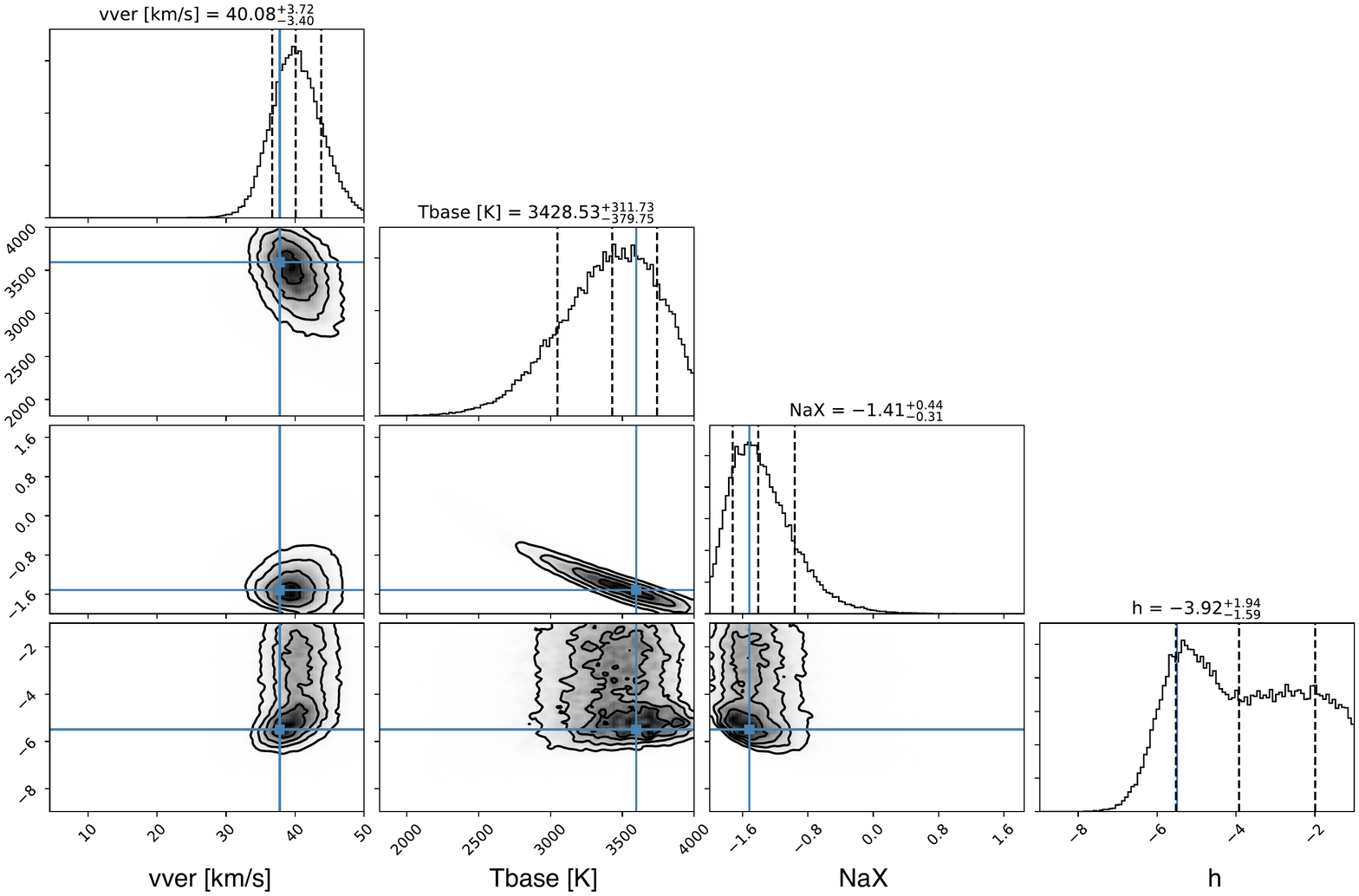}}
	\caption{Posterior distribution of isothermal line retrieval with an added vertical wind in the upper atmosphere. The parameter on the right $h$ is then used to calculate the pressure where the vertical velocity is added via Equation \ref{eq:pswitch}. The fraction of the atmosphere below that pressure has no winds. The distribution of $h$ shows that the retrieval shows a clear preference for high vertical winds for pressures below $h=-6$, but is unable to discriminate between the wind patterns in the lower parts of the atmosphere.}
	\label{fig:no_verposterior}
\end{figure*}

\begin{figure*}[hbt]
\resizebox{\textwidth}{!}{\includegraphics[trim=-1.0cm 9.0cm -1.0cm 9.0cm]{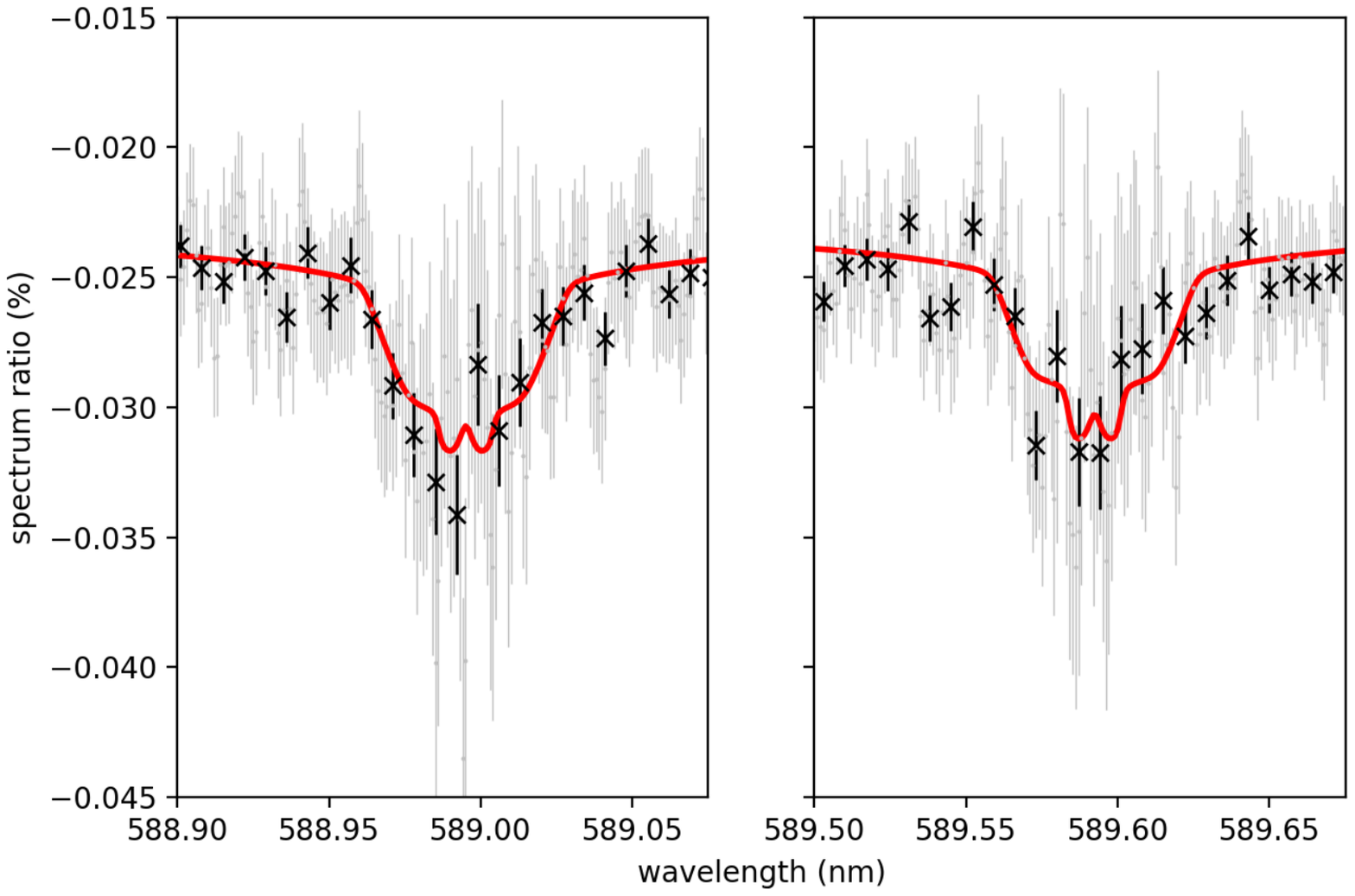}}
	\caption{Transmission spectrum of HD~189733b for both lines of the sodium doublet in grey, binned in black, with the best-fit retrieved by MERC in red. The best-fit was generated for the temperature gradient model with no added winds in the lower atmosphere and vertical winds in the upper atmosphere with its best-fit parameters from the posterior distribution in Figure \ref{fig:no_verposterior}.}
	\label{fig:nolowervverbest}
\end{figure*}

\end{appendix}

\end{document}